%% file: main.tex
\documentclass[10pt,journal,compsoc]{IEEEtran} 

\input{custom_config.sty}
\input{config.sty}

\begin{document}

\title{Why do Machine Learning Notebooks Crash? \\ {\LARGE An Empirical Study on Public Python Jupyter Notebooks}}

\author{
    Yiran~Wang~\orcidicon{0009-0007-4613-8960},~
    Willem~Meijer~\orcidicon{0000-0001-8482-3917},~
    José~Antonio~Hernández~López~\orcidicon{0000-0003-2439-2136},\\
    Ulf~Nilsson~\orcidicon{0000-0003-0269-9268}
    and~Dániel~Varró~\orcidicon{0000-0002-8790-252X},~\IEEEmembership{Senior Member,~IEEE}
    
    \IEEEcompsocitemizethanks{
        \IEEEcompsocthanksitem Yiran~Wang, Willem~Meijer and Ulf~Nilsson are with the Department of Computer and Information Science, Linköping University, SE-58183 Linköping, Sweden. 
        E-mail: yiran.wang@liu.se, willem.meijer@liu.se, ulf.nilsson@liu.se
         \IEEEcompsocthanksitem José~Antonio~Hernández~López is with the Department of Computer Science and Systems, University of Murcia, ES-30100 Murcia, Spain. 
        E-mail: joseantonio.hernandez6@um.es
        \IEEEcompsocthanksitem Dániel Varró is with the Department of Computer and Information Science, Linköping University, Sweden, McGill University, Canada, and Budapest Univ. of Technology and Economics, Hungary.
        E-mail: daniel.varro@liu.se
    }

}

%


\IEEEtitleabstractindextext{%
\begin{abstract}
Jupyter notebooks have become central in data science, integrating code, text and output in a flexible environment. With the rise of machine learning (ML), notebooks are increasingly used for prototyping and data analysis. However, due to their dependence on complex ML libraries and the flexible notebook semantics that allow cells to be run in any order, notebooks are susceptible to software bugs that may lead to program crashes. This paper presents a comprehensive empirical study focusing on crashes in publicly available Python ML notebooks. We collect 64,031 notebooks containing 92,542 crashes from GitHub and Kaggle, and manually analyze a sample of 746 crashes across various aspects, including crash types and root causes. 
Our analysis identifies unique ML-specific crash types, such as tensor shape mismatches and dataset value errors that violate API constraints.
Additionally, we highlight unique root causes tied to notebook semantics, including out-of-order execution and residual errors from previous cells, which have been largely overlooked in prior research.
Furthermore, we identify the most error-prone ML libraries, and analyze crash distribution across ML pipeline stages.
We find that over 40\% of crashes stem from API misuse and notebook-specific issues. 
Crashes frequently occur when using ML libraries like TensorFlow/Keras and Torch. Additionally, over 70\% of the crashes occur during data preparation, model training, and evaluation or prediction stages of the ML pipeline, while data visualization errors tend to be unique to ML notebooks.
\end{abstract}

}

\maketitle

\IEEEdisplaynontitleabstractindextext

%
\IEEEpeerreviewmaketitle

\newcommand{\rqone}{What are the predominant crash types and root causes of crashes in ML notebooks?}
\newcommand{\rqtwo}{To what extent are crashes in ML notebooks caused by ML bugs and by which ML libraries?}
\newcommand{\rqthree}{In which stages of the ML pipeline are crashes in notebooks most frequently to occur?}

\newcommand{\catone}{\textit{Crash type}}
\newcommand{\cattwo}{\textit{Root cause}}
\newcommand{\catthree}{\textit{ML/Python bug}}
\newcommand{\catfour}{\textit{ML pipeline}}
\newcommand{\catfive}{\textit{Library cause}}

\newcommand{\hreflib}[2]{\hrefcus{#1}{#2}}
\newcommand{\pandas}{\hreflib{https://pandas.pydata.org/}{pandas}}
\newcommand{\numpy}{\hreflib{https://numpy.org/}{numpy}}
\newcommand{\scipy}{\hreflib{https://scipy.org/}{scipy}}
\newcommand{\statsmodels}{\hreflib{https://www.statsmodels.org/stable/index.html}{statsmodels}}
\newcommand{\matplotlib}{\hreflib{https://matplotlib.org/}{matplotlib}}
\newcommand{\seaborn}{\hreflib{https://seaborn.pydata.org/}{seaborn}}
\newcommand{\plotly}{\hreflib{https://plotly.com/}{plotly}}
\newcommand{\missingno}{\hreflib{https://github.com/ResidentMario/missingno}{missingno}}
\newcommand{\wordcloud}{\hreflib{https://github.com/amueller/word_cloud}{wordcloud}}
\newcommand{\sklearn}{\hreflib{https://scikit-learn.org/stable/}{sklearn}}
\newcommand{\tensorflow}{\hreflib{https://www.tensorflow.org/}{\textbf{tensorflow}}}
\newcommand{\keras}{\hreflib{https://keras.io/}{\textbf{keras}}}
\newcommand{\torch}{\hreflib{https://pytorch.org/}{\textbf{torch}}}
\newcommand{\xgboost}{\hreflib{https://xgboost.readthedocs.io/en/stable/}{xgboost}}
\newcommand{\lightgbm}{\hreflib{https://lightgbm.readthedocs.io/en/stable/}{lightgbm}}
\newcommand{\catboost}{\hreflib{https://github.com/catboost/catboost}{catboost}}
\newcommand{\imblearn}{\hreflib{https://imbalanced-learn.org/stable/}{imblearn}}
\newcommand{\cvtwo}{\hreflib{https://opencv.org/}{cv2}}
\newcommand{\torchvision}{\hreflib{https://pytorch.org/vision/stable/index.html}{\textbf{torchvision}}}
\newcommand{\skimage}{\hreflib{https://scikit-image.org/}{skimage}}
\newcommand{\nltk}{\hreflib{https://www.nltk.org/}{nltk}}
\newcommand{\transformers}{\hreflib{https://huggingface.co/docs/transformers/}{\textbf{transformers}}}
\newcommand{\optuna}{\hreflib{https://optuna.org/}{optuna}}
\newcommand{\datasets}{\hreflib{https://huggingface.co/docs/datasets/}{datasets}}

\input{Content/introduction}

\input{Content/background}

\input{Content/relatedwork}

\input{Content/method}

\input{Content/results}

\input{Content/discussion}

\input{Content/validity}

\input{Content/conclusion}

\ifCLASSOPTIONcompsoc
  \section*{Acknowledgments}
\else
  \section*{Acknowledgment}
\fi

This work was partially supported by the Wallenberg AI, Autonomous Systems and Software Program (WASP) funded by the Knut and Alice Wallenberg Foundation, and the Software Center Project 61.

\ifCLASSOPTIONcaptionsoff
  \newpage
\fi

\bibliographystyle{IEEEtran}
\bibliography{References}

\end{document}

%% file: Content/introduction.tex
\IEEEraisesectionheading{\section{Introduction}\label{sec:introduction}}



\IEEEPARstart{J}{upyter} notebooks, such as those used in Kaggle, JupyterLab, and Google Colab, have become indispensable in data science due to their interactive nature~\cite{Wang20notebook}.
They offer real-time feedback during code execution, making them ideal for tasks such as data visualization, pattern exploration, and data analysis~\cite{Pimentel19notebook, Koenzen20notebook}. Notebooks also streamline documentation by combining text, code, execution results, and visualizations into a single, cohesive environment, designed to enhance collaboration and reproducibility~\cite{Pimentel19notebook, Wang20notebook}.

With the growing availability of data, machine learning (ML) and deep learning (DL) techniques have gained popularity for data modeling and decision-making within the broader field of data science~\cite{jahani2023data, sohail2023genetic}. 
In this context, Jupyter notebooks, particularly Python-based ones, have been widely adopted for ML prototyping and data analysis~\cite{Grotov22notebook}. 
As confirmed by several industrial collaborators~\cite{swc2024}, Jupyter notebooks often serve as the starting point for developing ML applications before transitioning to production-ready ML pipelines, where much of the Python code used in notebooks is reused in large-scale distributed computing environments.
This widespread adoption, even in mission-critical applications such as those at NASA Jet Propulsion Lab~\cite{nasahecc2024}, underscores the importance of quality assurance for notebooks. Bugs that arise during the notebook stage may persist into later development stages, potentially affecting deployed products. Therefore, studying notebook bugs is essential to improve the overall reliability and robustness of ML-driven software.

Software bugs in ML notebooks may be attributed to various underlying root causes.
Such bugs commonly stem from the heavy reliance on ML libraries used for various stages in a ML pipeline including data processing and ML model building or other data modeling algorithms~\cite{islamComprehensiveStudyDeep2019}. These libraries often come with steep learning curves and varying abstraction levels, leading to errors in API usage, configuration, and compatibility. 
Furthermore, the flexible semantics of Jupyter notebooks, such as allowing code cells to be executed out of order, could increase the likelihood of bugs~\cite{Pimentel19notebook, Wang20notebook,desantanaBugAnalysisJupyter2024}. Understanding these challenges is crucial for improving the debugging process and developing more robust tools for notebook environments. 
However, most previous studies~\cite{zhangEmpiricalStudyTensorFlow2018, islamComprehensiveStudyDeep2019, humbatovaTaxonomyRealFaults2020, zhangEmpiricalStudyProgram2020, Morovati24bug} focus on ML scripts, which are standalone, linear code files (\eg~.py files), rather than ML programs developed in notebooks. In this paper, we refer to ML programs as a broader category that includes both ML scripts and ML notebooks.

Bugs in ML programs can manifest in multiple ways, such as low effectiveness, reduced efficiency, or \textit{crashes}~\cite{zhangEmpiricalStudyTensorFlow2018} which abruptly terminate the execution. In Python, crashes occur as unhandled exceptions, representing the most severe bug symptom as execution is halted and needs immediate attention.
Existing research indicates that crashes are the most common symptoms of bugs in ML programs~\cite{islamComprehensiveStudyDeep2019, Morovati24bug, desantanaBugAnalysisJupyter2024}. Crashes are easier to investigate due to accompanying error information that can be captured in the \textit{notebook} environment.

We conduct an empirical study to analyze 92,542 crashes in 64,031 publicly available Jupyter notebooks for ML in Python (referred to as \emph{ML notebooks} throughout the paper). We identify prevalent crash types and root causes, and investigate which ML libraries and pipeline stages most frequently contribute to crashes.
Furthermore, we highlight root causes related to the improper use of notebook semantics (\eg~out-of-order execution) which has not been covered in prior research.
Our long-term goal is to enhance the use of Jupyter notebooks for ML development, guide the creation of better debugging tools, and improve the overall quality of ML notebooks. To our best knowledge, this is the first study focused on analyzing crashes in ML notebooks.

This study incorporates ML notebooks, used in Jupyter-based environments like Kaggle and JupyterLab, which share the same execution model and \texttt{.ipynb} format.
We examine crashes in notebooks collected from representative repositories such as GitHub and Kaggle, which serve distinct user groups with different coding practices. Many data scientists and ML practitioners, particularly in exploratory and collaborative workflows, may lack strong software engineering skills and often commit notebooks with unresolved bugs~\cite{desantanaBugAnalysisJupyter2024}. By focusing on those notebooks containing crash information, we gain insights into real-world debugging challenges and persistent errors in ML development, which are addressed along three main research questions. We also analyze how crash characteristics vary between GitHub and Kaggle, providing a crosscutting aspect of our analysis.

\myrq{1} {\rqone}
{
Crashes in notebooks often provide detailed error information including exception types that describe the direct trigger of the crash. The root cause, on the other hand, is the underlying reason behind a crash such as wrong usage of APIs, mismatched data expectations, or improper use of notebook semantics (\eg~out-of-order execution).
Identifying the most common crash types, root causes, and their relationship helps developers and researchers prioritize efforts to handle the most frequent issues in a systematic way.
This makes the development of ML applications in notebooks more efficient and productive, while developers can still benefit from the flexibility of Jupyter notebooks.
}
{
In our study, the most common crash types in ML notebooks are \texttt{Variable Not Found} (variable name not found), \texttt{Invalid Argument} (function receives an inappropriate argument), and \texttt{IO Error} (errors when handling files), attributing to 37.3\% of crashes in total. 
We identify ML-specific crash types including \texttt{Tensor Shape Mismatch}, \texttt{Data Value Violation}, \texttt{Out of Memory}, \texttt{Unsupported Broadcast}, \texttt{Feature Name Mismatch}, and \texttt{Model Initialization Error}.
The main root causes include API misuse (20.9\%) and notebook-specific issues (19.4\%, mostly due to out-of-order execution). 
Our analysis reveals that name-related errors (\ie~undefined names) are primarily caused by notebook-specific issues. API misuse, implementation error, and data confusion are the primary causes of a wide range of crash types.
}

\myrq{2} {\rqtwo}
{
Crashes in ML notebooks can result from ML-related bugs and general Python bugs. We examine to what extent these crashes are caused by ML bugs, including those that arise from code using ML APIs or involving variables, objects, or other components derived from ML APIs.
Additionally, we determine whether a crash results from the direct use of an ML library to identify which libraries are more often associated to crashes and pose greater debugging challenges.
This helps guide better ML API usage practices and improves ML library stability.}
{
In our study, ML bugs account for 67.7\% of ML notebook crashes, with 62.8\% on GitHub and a significantly higher proportion on Kaggle (73.0\%). The primary causes of ML-related crashes are API misuse (25\%), data confusion (20\%), and notebook-specific issues (20\%), while general Python bugs are mostly due to implementation (29\%) and environment (28\%) errors. API misuse is more common in ML-related crashes compared to Python crashes, highlighting the increased difficulty of using ML libraries correctly.
In particular, \texttt{tensorflow/keras} and~\texttt{torch} are shown to be the most common source of crashes with the highest crash ratio relative to its usage.
}

\myrq{3} {\rqthree}
{
Studying which stages of the ML pipeline are most prone to crashes, such as data preparation, model construction, or model training stages, helps developers and researchers focus on improving robustness in critical stages, and optimize development resources accordingly.
This does not imply that all ML notebooks follow a complete pipeline. For example, some may exclusively focus on data preparation, omitting other stages such as model construction or training. Therefore, we analyze which stages of a general ML pipeline caused the crash, regardless of whether a notebook implements all of them.
}
{
In our study, crashes predominantly occur during data preparation (33.0\%), model training (19.4\%), and evaluation/prediction (18.4\%) stages. GitHub has more crashes during environment setup and data preparation, whereas more crashes are observed during the evaluation/prediction stage in Kaggle. Additionally, over 70\% of the crashes during model training and evaluation/prediction are caused by ML bugs.
}

\noindent\textbf{Replication package.} All experimentation artifacts can be found on GitHub~\cite{wang2025code}, and data on Zenodo~\cite{wang2025data}.


%% file: Content/background.tex
\section{Background and related work}
\label{sec: background}

In this section, we describe the concept of a Jupyter notebook and explain the notations used throughout the paper. We then review empirical studies related to bugs in ML and data science programs. 

\input{Content/Subcontents/fig_motivating_example}
\input{Content/Subcontents/tb_relatedwork}

\subsection{Terminology and motivating example}

First, we clarify some key terms following the definitions provided in a previous study~\cite{avizienis2004faultdef}. We use \textit{bugs} or \textit{faults} as inherent defects that result in incorrect or unexpected behaviors, known as \textit{errors}.  If left unaddressed, these errors can lead to \textit{failures} or \textit{crashes}. In Python, crashes manifest as \textit{exceptions}, typically accompanied by error information.

Jupyter notebooks are the most common medium of ML prototyping and data analysis~\cite{Grotov22notebook}. Notebooks are more flexible compared to, for example, Python scripts by offering additional interactive features. \autoref{fig: motivating_sample} shows an example excerpt of an ML notebook that trains and evaluates a neural network. A \emph{notebook} typically consists of multiple \emph{cells}, which can contain either documentation or executable code. When a user runs a code cell, Jupyter uses a kernel to execute it. After execution, the notebook assigns an \emph{execution number} to the cell, indicating the order in which it was run. The \emph{output} (if any) is displayed directly below the code cell. In the example, two code cells have been executed, their outputs are shown below, and the execution numbers appear to the left of each cell, marking the execution order.

When an exception occurs during the execution of a code cell (\ie~the code crashes), Jupyter provides detailed error information in the output.
The error information includes three key components: \emph{exception type}, \emph{error value}, and a \emph{traceback} that records the sequence of function calls leading to the crash. Analyzing this error information is crucial because it offers direct, first-hand insights into the exact moment and nature of the crash, helping developers understand the immediate cause of failure. 

The unique flexibility of Jupyter notebooks introduces additional challenges that may affect how crashes occur. Notebooks allow users to execute cells in any order, run the same cell multiple times, and skip executing cells. While this supports iterative and exploratory development, it can lead to unexpected behavior if cells are executed out of sequence. For example, the notebook in \autoref{fig: motivating_sample} can be executed in reverse, such that the code that evaluates the model is executed before the code that compiles and trains the model. In our example, this would cause a crash.

%% file: Content/Subcontents/fig_motivating_example.tex
\begin{figure}[b]
\centering
\includegraphics[width=0.9\linewidth]{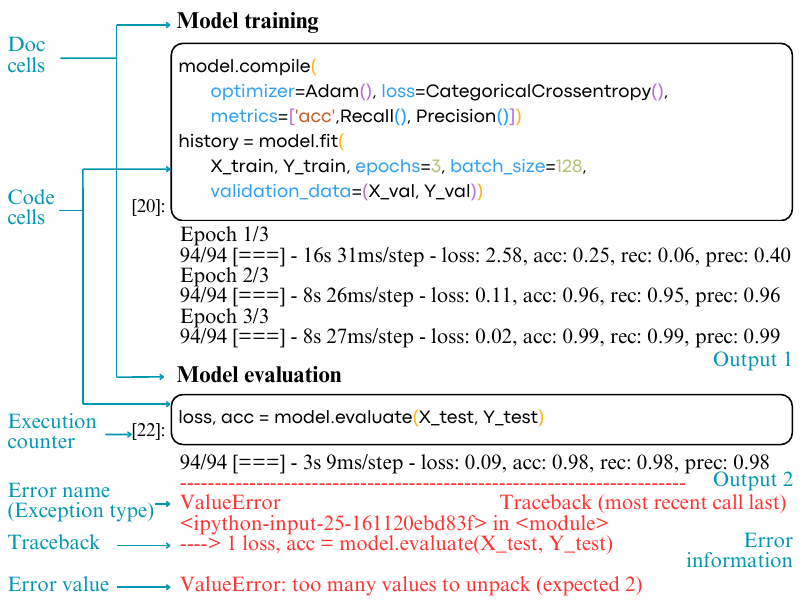}
\caption{A real Jupyter notebook that trains and evaluates a neural network with documentation (doc cells), Python code (code cells), and execution logs (output and a crash with error information).}
\label{fig: motivating_sample}
\end{figure}

%% file: Content/Subcontents/tb_relatedwork.tex
\begin{table*}[ht]
    \scriptsize
    \centering
    \caption{Related work of empirical studies on bugs in ML, DL or data science programs, including the considered analysis dimensions, software artifacts, targeted ML libraries, and used data sources: GitHub (GH) and Stack Overflow (SO).}
    \begin{tabular}{m{3.3cm}|m{3.3cm}|m{2.9cm}|m{3cm}|m{3.4cm}}
        \hline
        \thead{Study} & \thead{Analysis dimensions} & \thead{Targets} & \thead{Libraries} & \thead{Data sources}\\
        \hline
        Zhang~\etal~\cite{zhangEmpiricalStudyTensorFlow2018} 
        &
        Root cause, symptom
        &
        DL scripts
        &
        TensorFlow
        &
        SO posts, GH issues\\
        
        \hline
        
        Islam~\etal~\cite{islamComprehensiveStudyDeep2019} 
        & 
        Bug type, root cause, symptom, ML pipeline
        & 
        DL scripts
        &
        Caffe, Keras, TensorFlow, Theano, PyTorch
        &
        SO posts, GH commits\\
        
        \hline
        
        Humbatova~\etal~\cite{humbatovaTaxonomyRealFaults2020}
        & 
        Fault taxonomy (root cause)
        &
        DL scripts
        &
        TensorFlow, Keras, PyTorch
        &
        SO posts, GH commits/issues, interviews, survey\\
        
        \hline
        
        Zhang~\etal~\cite{zhangEmpiricalStudyProgram2020}
        & 
        Failure type, root cause, ML pipeline
        &
        DL scripts
        &
        Not limited
        &
        Failure messages from a Microsoft DL platform, interviews\\
        
        \hline
        
        Morovati~\etal~\cite{Morovati24bug} & 
        ML/non-ML bugs, root cause and symptom of ML bugs
        &
        ML(DL) systems
        &
        TensorFlow, Keras, PyTorch
        &
        GH issues\\
        
        \hline

        Ahmed~\etal~\cite{ahmedCharacterizingBugsPython2023} & 
        Bug type, root cause, symptom
        &
        Data analysis programs
        &
        Most tagged ones on SO
        &
        SO posts, GH commits/issues\\
        \hline
        
        Thung~\etal~\cite{thung12bugml}, Sun~\etal~\cite{sunEmpiricalStudyReal2017}, Jia~\etal~\cite{jiaEmpiricalStudyBugs2020}, Chen~\etal~\cite{chenUnderstandingDeepLearning2023} 
        & 
        Bug type, root cause, symptom, severity
        &
        ML frameworks
        &
        OpenNLP, Paddle, Caffe, TensorFlow, PyTorch, etc
        &
        GH repositories, bug reports (\eg~JIRA)\\

        \hline

        De Santana~\etal~\cite{desantanaBugAnalysisJupyter2024} 
        & 
        Bug type, root cause, symptom
        &
        Data science Jupyter notebooks
        &
        Not limited
        &
        SO posts, GH commits, interviews, survey\\
        
        \hline
        
        Our study & 
        Exception type, root cause, ML/Python bug, ML pipeline
        &
        ML Jupyter notebooks
        &
        Not limited
        &
        Error outputs from GH and Kaggle Jupyter notebooks\\
        \hline
        
    \end{tabular}
    \label{tb: relatedwork}
\end{table*}

%% file: Content/relatedwork.tex
\subsection{Studies of bugs in ML or data science programs}

\autoref{tb: relatedwork} summarizes prior empirical research on bugs in ML, DL, and data science programs. It highlights the analysis dimensions, software artifacts, targeted ML libraries, and data sources used in these studies.

The majority of prior research~\cite{zhangEmpiricalStudyTensorFlow2018, islamComprehensiveStudyDeep2019, humbatovaTaxonomyRealFaults2020, zhangEmpiricalStudyProgram2020, Morovati24bug} focuses on bugs in ML/DL programs written in \emph{scripts}. Other studies have examined bugs in popular ML frameworks~\cite{thung12bugml, sunEmpiricalStudyReal2017, jiaEmpiricalStudyBugs2020, chenUnderstandingDeepLearning2023} or data analysis programs~\cite{ahmedCharacterizingBugsPython2023}. These works have provided valuable insights into the nature of bugs in ML code, helping developers understand common error patterns. The identified root causes include incorrect model parameters or structures, API misuse,  and unaligned tensors. Morovati~\etal~\cite{Morovati24bug} take a different perspective of software maintenance by comparing ML bugs with non-ML bugs, broadening the understanding of bugs in ML/DL systems.

While these studies provide a strong foundation, very few have specifically targeted Jupyter notebooks. One notable exception is the work by De Santana~\etal~\cite{desantanaBugAnalysisJupyter2024}, which classifies bugs into eight categories, concentrating on issues related to environment setup, kernel, notebook conversion, and implementation issues. They explore root causes such as configuration and version conflicts, and coding errors, examining their impacts. However, their study does not take into account the iterative and flexible nature of Jupyter notebooks. Furthermore, it primarily examines general bugs without providing an in-depth analysis of those specific to ML. Moreover, no studies focus on program crashes. 

Our study addresses these gaps by analyzing crashes in Python ML notebooks with investigation of how notebook semantics (\eg~out-of-order execution, or previous cell error) impact crashes. Although prior work~\cite{Pimentel19notebook} has linked out-of-order execution to reproducibility issues, we are the first to identify notebook-specific semantic issues as root causes of crashes. By examining error information in cell outputs, we distinguish \textit{ML bugs} from general Python bugs and pinpoint the stages of the \textit{ML pipeline} where these crashes occur.

Most prior research~\cite{zhangEmpiricalStudyTensorFlow2018, islamComprehensiveStudyDeep2019, humbatovaTaxonomyRealFaults2020, Morovati24bug, thung12bugml, sunEmpiricalStudyReal2017, jiaEmpiricalStudyBugs2020, chenUnderstandingDeepLearning2023, ahmedCharacterizingBugsPython2023, desantanaBugAnalysisJupyter2024} draws from data sources such as Stack Overflow posts, GitHub issues, or code commits.
While these sources offer valuable insights, they often describe bugs using natural language, which can abstract away important details about the actual execution context, making it harder to generalize findings. Additionally, Zhang~\etal~\cite{zhangEmpiricalStudyProgram2020} examine DL job failures from an industrial platform. However, job failure messages typically capture issues only after job submission, potentially overlooking critical bugs that occur during coding. In contrast, crash information in Jupyter notebooks provides real-time feedback tied directly to code execution, offering precise and context-rich data. This allows us to capture bugs as they occur, providing more accurate insights into the nature of crashes. By analyzing this crash information, we can identify and diagnose issues more effectively.

Furthermore, many previous studies focused on bugs in ML/DL programs that use specific ML libraries~\cite{zhangEmpiricalStudyTensorFlow2018, islamComprehensiveStudyDeep2019, humbatovaTaxonomyRealFaults2020, ahmedCharacterizingBugsPython2023, Morovati24bug}, or defined ML bugs as all errors that occur in ML components in a system~\cite{Morovati24bug}. In contrast, we examine a broad range of code from our dataset and identify popular ML libraries. As we investigate ML bugs which cause crashes of code that interacts with such libraries, our approach provides a more direct link between crashes and ML libraries.


%% file: Content/method.tex
\input{Content/Subcontents/fig_data_process}

\section{Experimental setup} \label{sect: method}

This section outlines the experimental setup of our study, as shown in \autoref{fig: dataset}. We begin with the data collection methods, followed by the filtering criteria for identifying the relevant population. Next, we describe the sampling strategy used to select a representative sample. Lastly, we describe the process for manual analysis and resampling.

\subsection{Data sources and mining method} \label{subsect: data_collect}

Our data consists of Jupyter notebooks from GitHub and Kaggle, containing source code, documentation, and output cells (including error information). \emph{GitHub} is a widely used platform for open-source projects, providing notebooks from a broad community of developers. \emph{Kaggle} is known for data science and ML competitions, offering notebooks from more specialized ML practitioners. By including both platforms, we capture diverse user bases and practices, allowing us to explore potential differences in their crash behaviors.

GitHub notebooks are downloaded from the deduplicated \emph{The Stack} dataset~\cite{Kocetkov2022TheStack}, which includes active GitHub repositories from January 1, 2015, to March 31, 2022. Kaggle notebooks are retrieved using the KGTorrent database~\cite{KGTorrent_Quaranta}, based on the latest (February 7, 2024) Meta Kaggle data~\cite{megan2022metakaggle}. Meta Kaggle is an official and daily updated collection of public metadata of Kaggle notebooks. 
We download Kaggle Python notebooks with outputs between January 1, 2023, and January 1, 2024. We limit to this time frame because Kaggle limits the number of daily HTTP requests.
The time gap between GitHub and Kaggle also indicates fewer overlapping notebooks. This results in a total of 1,070,293 GitHub notebooks and 166,655 Kaggle notebooks.

\subsection{Data filtering} \label{subsect: data_filtering}

We filter the notebooks to identify valid ML notebooks and meaningful crashes, ensuring data relevance and quality. 

\paragraph*{\textit{Python notebooks}}
This study focuses on Python notebooks. To identify the programming language of each notebook, we use the \textit{Guesslang} library~\cite{guesslang2024} used in previous works~\cite{zhang2021study,gong2024cosqa} for similar filtering steps. The top three most commonly used languages in GitHub notebooks are Python (95.35\%), Julia (1.22\%), and R (0.92\%). We then filter out non-Python notebooks, leaving a total of 1,020,540 Python GitHub notebooks. All Kaggle notebooks are written in Python.

\paragraph*{\textit{Notebooks that contain crashes}}
We aim to study notebook crashes. Hence, we parse and filter all notebooks to retain only those that contain crashes in their cell outputs. 
Specifically, we filter notebooks containing at least one cell such that 1) its \texttt{cell\_type} is \texttt{code}, 2) its \texttt{output\_type} is \texttt{error}. 
After applying this filter, we observe that 10.67\% of the collected GitHub notebooks and only 2.61\% of Kaggle notebooks contain crashes. 
This disparity may stem from Kaggle’s role as a platform that encourages developers to share complete, correct, and well-documented notebooks, allowing others to learn from real-world data science challenges and solutions. In total, there are 108,925 GitHub and 4,349 Kaggle notebooks containing 172,638 and 6,745 crashes (\ie~error outputs) respectively.


\paragraph*{\textit{Valid notebooks}} \label{para: data_filtering_valid}
As our study targets runtime crashes during execution, we exclude notebooks with syntactically incorrect code. Syntax errors prevent execution, making the notebook non-reproducible and unreliable (garbage-in, garbage-out). Furthermore, we investigate notebook-specific issues, such as previous cell errors, focusing on previous cells that execute incorrectly rather than contain syntactically invalid code. 
Including crashes caused by invalid code would highlight problems that can be resolved using basic static analysis tools, thus yielding no insights into the unique challenges faced during ML notebook development.
Additionally, notebooks with persistent syntax errors are typically of lower quality, often containing incomplete work, and may not reflect typical ML development workflows.
To check for syntactic validity, we use Python 3 Abstract Syntax Tree  (AST) module, which parses code according to Python 3 grammar. Since Python 3 is the current standard, this approach also filters outdated notebooks, aligning with modern ML workflows.
 
\paragraph*{\textit{ML notebooks}}
Jupyter notebooks can be used for various applications. This paper focuses specifically on ML notebooks, defined as those that use at least one ML library. To compile a comprehensive list of ML libraries, we rely on Kaggle as a trusted source, given its reputation as a prominent platform for ML or data science competitions with an active online community. 
We extract and rank all libraries imported by Kaggle notebooks based on frequency.
The top 50 libraries, used in $\sim$95\% of all Kaggle notebooks, are independently reviewed by two authors to determine their relevance to ML. Full agreement is reached in identifying 24 popular ML libraries (see \autoref{tb: top_ML_libs}). We then use these libraries to identify ML notebooks in our dataset.

\input{Content/Subcontents/tb_top_libs}

\paragraph*{\textit{KeyboardInterrupt exceptions}} \label{para: data_filtering_keyboard}
A \textit{KeyboardInterrupt} exception happens when a user manually stops program execution (\eg~with Ctrl+C). We exclude notebooks containing only \texttt{KeyboardInterrupt} type of exceptions because it is user-triggered and does not necessarily reflect code issues.

\paragraph*{\textit{Enriching and filtering non-descriptive error values}}
Normally, error values of crashes can be extracted from notebooks without extra processing. But, in some cases, developers substitute error values with non-descriptive messages like “ignored” or “null”. We replace them with a more descriptive message extracted from the traceback of the crash. When this is not possible, the respective crash is removed from the dataset.

The final population comprises 61,342 GitHub notebooks with 88,667 crashes and 2,689 Kaggle notebooks with 3,875 crashes, in total 64,031 notebooks with 92,542 crashes.

\subsection{Sampling} \label{subsec: cluster_sample}

Our study requires a manual labeling and review process to analyze ML notebook crashes. Since investigating the entire population is infeasible, we select a meaningful sample of crashes for manual analysis.

We apply \textit{proportional stratified sampling}~\cite{CHAN1996775sampling}, a strategy that builds on top of simple random sampling. This method divides the population into strata and randomly samples from each group in proportion to its size. We base these strata on error values as they capture more detailed information of crashes compared to, for example, exception types alone. Additionally, crashes with different exception types but similar error values often exhibit similarities, suggesting they share common root causes or other patterns. 

\paragraph*{\textit{Clustering crashes}}
To group similar error values of crashes, we apply a clustering method.
We first preprocess error values to generalize them, such as removing unnecessary substrings like URLs, file paths, function names, variable names, and punctuation. This makes the error values more representative of common crash patterns rather than overly specific to individual crashes. 

We then cluster the preprocessed error values using Jaccard similarity~\cite{tan2005datamining} which measures overlap between sets. This method is ideal for grouping error values with similar patterns, as shown in previous studies~\cite{wang2020log, allamaniscodedup2019}. An example of clustering results using Jaccard similarity is presented in~\autoref{tb: ja_sim}.
We choose a 0.7 threshold for Jaccard similarity, based on its use in a previous study~\cite{allamaniscodedup2019} in the context of code clone detection. We also test smaller and larger thresholds (ranging from 0.5 to 0.9 in 0.1 intervals). After evaluating 50 clusters formed at each threshold, we find that the smaller thresholds lead to clusters with mixed error messages, while the larger ones create too many similar clusters. The 0.7 threshold provides a coherent and meaningful grouping, making it the optimal choice for our analysis.
We first cluster all crashes (both GitHub and Kaggle) to get a comprehensive view of common patterns, then calculate clusters for GitHub and Kaggle datasets separately to identify platform-specific differences in crash behaviors. Each pair of error values is compared, and if their similarity exceeds the threshold, they are connected. Crashes are grouped into the same cluster if they can be reached through any connected crashes, ensuring reachability within the cluster. We obtain a total of 7,433 clusters. Separate statistics for each platform reveal 7,222 clusters for GitHub and 640 for Kaggle.

\input{Content/Subcontents/tb_jaccard_sim}
\input{Content/Subcontents/tb_cluster_exp}
\input{Content/Subcontents/tb_dimensions}

We further visualize the clusters in relation to exception types. We observe that one exception type may have different patterns that lead to different crash types as~\autoref{tb: cluster_exp} shows.
This is likely because Python exceptions can be customized and raised by users or library developers under various names or reasons. This finding motivates our approach of clustering based on error values, rather than using exception types directly. 

We include all clusters, emphasizing the most frequent ones, which cover over 70\% of crashes in each dataset - 65 clusters for GitHub with a minimum size of 100 and 50 for Kaggle with a minimum size of 10. The remaining, less common crashes are grouped into a single cluster. This results in 66 clusters for GitHub and 51 for Kaggle overall.

\paragraph*{\textit{Proportional-to-cluster-size sampling}}
We estimate sample sizes using random sampling without replacement~\cite{lohr2021sampling, tort1978sampling}. We calculate sample sizes and analyze sampled crashes separately for GitHub and Kaggle because they may exhibit different behaviors. The desired sample sizes are determined with a 95\% confidence interval (CI) and a 5\% margin of error (MoE). 
This results in sample sizes of 360 for GitHub and 332 for Kaggle crashes. 
The details of the sample size calculation process can be found in our GitHub repository~\cite{wang2025code}.

We then use these sample sizes and the identified clusters (GitHub: 66, Kaggle: 51) to conduct proportional sampling based on cluster size. To ensure comprehensive coverage, we sample at least one crash from each cluster, as recommended by Chan~\etal~\cite{CHAN1996775sampling}. This yields a total of 390 sampled crashes from GitHub and 356 from Kaggle.

\subsection{Crash analysis} \label{subsec: analysis_dim}

We manually analyze crashes from various perspectives. We establish a set of \emph{dimensions} to describe crashes in ML notebooks. \autoref{tb: dimensions} shows the 4 dimensions considered in our analysis and how they relate to our research questions.


\paragraph*{{\MakeUppercase{\textbf{\catone}}}}
The first dimension categorizes crashes based on the exceptions that trigger them, originally shown as the error names in the notebook outputs (see~\autoref{fig: motivating_sample}). This classification is directly linked to the specific runtime exceptions responsible for a crash. A comparable approach is used by Zhang~\etal~\cite{zhangEmpiricalStudyProgram2020}, who classified job failures, which are instances where the job crashes, according to their runtime error types.

We manually refine the raw exception types for two reasons. First, we distinguish between Python's built-in exceptions and those defined by library developers or users. Our analysis reveals that 7.21\% of crashes on GitHub and 7.31\% on Kaggle have names differing from Python’s built-in types. By refining exception types, we can better identify customized exceptions. Moreover, as shown in \autoref{tb: cluster_exp}, one exception type may exhibit different patterns. Users can define and raise exception types freely, which may not always accurately represent the true trigger of a crash. Therefore, manually refining these types allows us to enhance our analysis of the crashes.


\paragraph*{{\MakeUppercase{\textbf{\cattwo}}}}
This dimension categorizes root causes of crashes, focusing on the fundamental issues that lead to these crashes. Unlike traditional software, ML programs face unique challenges, such as complex data processing, model construction, and extensive use of ML APIs. The flexibility of notebooks, which allows cells to be executed at any time, can introduce further complications.

We base our initial root cause categories on prior studies (API misuse~\cite{islamComprehensiveStudyDeep2019, zhangEmpiricalStudyProgram2020}, data confusion~\cite{zhangEmpiricalStudyProgram2020}, implementation error~\cite{zhangEmpiricalStudyProgram2020, desantanaBugAnalysisJupyter2024}, library issues~\cite{islamComprehensiveStudyDeep2019}, environment issue~\cite{desantanaBugAnalysisJupyter2024}, insufficient resource~\cite{desantanaBugAnalysisJupyter2024}). We extend this initial set of categories based on further insights from our pre-study. We discover how notebook semantics, such as out-of-order execution and previous unresolved errors, contribute to crashes. Additionally, we identify ML model mismatches as a root cause, occurring when a previously saved model is incompatible with the ML model defined in the current code.
By categorizing crashes by root causes, we capture a broad range of crash sources, from misunderstanding of data or APIs to environment and notebook-specific issues, obtaining a comprehensive understanding of why the crashes happen.

\paragraph*{{\MakeUppercase{\textbf{\catthree}}}}
This dimension classifies crashes based on whether they occur in code related to ML libraries. Morovati~\etal~\cite{Morovati24bug} compare ML bugs and non-ML bugs by identifying ML bugs as all bugs within ML components of DL systems. In contrast, our study focuses on the ML components themselves, aiming to evaluate how often crashes stem from interactions with ML libraries. We define \textit{ML bugs} as crashes that occur in code using ML libraries or objects derived from them. By distinguishing ML bugs and general Python bugs, we can better understand the challenges unique to ML development, including the use of ML APIs, model handling, and data processing tasks.

\paragraph*{{\MakeUppercase{\textbf{\catfive}}}}
This dimension labels which ML library is directly responsible for the crash. In our ML/Python bug classification, a crash is considered an ML bug if the code involves any ML APIs or components, regardless of whether the issue directly stems from ML libraries. This dimension focuses specifically on crashes caused by ML libraries, meaning that the crash must occur as a direct use of the library. As such, this is a subset of ML bugs. This dimension can contribute to determine which ML libraries are more challenging to use.

\paragraph*{{\MakeUppercase{\textbf{\catfour}}}}
In this dimension, we categorize crashes based on where they occur during the \catfour~to identify critical stages.
We define key stages building on the framework from a prior study by Zhang~\etal~\cite{zhangEmpiricalStudyProgram2020} (environment setup, data preparation, training, and evaluation/prediction stages) and a pre-study of our dataset (data visualization and model construction stages).
This categorization does not imply that all notebooks adhere to a strict pipeline. For example, some only implement a subset of stages, while others contain the same stage multiple times (\eg~when they try and compare different preprocessing methods). Rather, it provides a structured way to analyze where crashes occur within a general ML workflow and where crashes in the ML pipeline are more frequent and potentially harder to resolve.


\subsection{Manual labeling process} \label{subsec: manual_label}
To classify ML notebook crashes, our process is structured in two stages: labeling and reviewing.

In the \emph{labeling} stage, we categorize crashes across the predefined dimensions introduced in \autoref{subsec: analysis_dim}. We adopt grounded theory methods~\cite{seaman1999quality}, incorporating first-cycle and second-cycle coding phases~\cite{saldana2015coding}.

We first label crashes based on predefined initial categories for each dimension (see~\autoref{subsec: analysis_dim}). We then iteratively refine these categories whenever new crash patterns emerge that do not fit within the existing ones, particularly for \catone~and \cattwo. We do this by adding new categories and updating existing ones, which is common in grounded theory~\cite{seaman1999quality}. For example, the crash type \texttt{Tensor Shape Mismatch} was originally included in \texttt{Value Error} and later split into a separate category. This creates more consistent and well-separated categories.

During the first-cycle coding, three evaluators independently label an initial set of 50 crashes from the sampled dataset. Following this, the evaluators meet to synchronize their findings, discuss the results, and establish uniform labeling principles. This collaborative discussion promotes consistency and mitigates potential bias early in the process. To further improve consistency, we organize crashes from both GitHub and Kaggle datasets by clusters, which groups similar crashes together for labeling and review. Additionally, evaluators reference all notebooks containing the sampled crashes to access relevant crash contexts. This structured approach, grounded in an open coding scheme~\cite{seaman1999quality}, allows category expansion while minimizing bias and enhancing consistency in our classifications.
During the second-cycle coding, the remaining crashes are labeled independently. Regular meetings are held to discuss and reach a consensus on any disagreements, promoting consistent standards across the entire dataset.

In the \emph{reviewing} stage, a separate evaluator validates the labeling results. In cases of disagreement, discussions take place to resolve conflicts, and if needed, all three evaluators come together to reach a final decision. 

\input{Content/Subcontents/tb_stats_manual_label}

\subsection{Resampling} \label{subsec: sample_resample}

After manually labeling and reviewing the data, we identify 123 cases in which the crashes do not fit into any stage of the ML pipeline. The majority of these cases originate from tutorial notebooks, educational materials, or physics simulations. Additionally, we find four irrelevant crashes (\eg~cells containing random text, exceptions raised to print comments). As these instances do not represent actual crashes in ML development, we exclude them from our analysis.


To maintain statistically significant sample sizes, we use a resampling approach similar to proportional-to-cluster-size sampling (\autoref{subsec: cluster_sample}), where replacements are drawn from the same clusters where exclusions occur. We then verify whether the replacements should be excluded as well. If such cases exist, we iteratively perform resampling until we reach the target sample sizes (GitHub: 390, Kaggle: 356). Afterwards, we manually label the 127 resampled crashes following the same process (see \autoref{subsec: manual_label}).


The change rates before and after reviewing are shown in \autoref{tb: stats_manual_label}. The reviewing process sometimes results in updates to earlier categories. For example, many changes are made to categorize crashes in~\catfour~based on the stage where the code is written, rather than by its functionality, which leads to a high rate of reclassification. On the other hand, the high change rate of~\cattwo~reflects the challenges of accurately identifying the root causes of the crashes.

%% file: Content/Subcontents/fig_data_process.tex
\begin{figure}
\centering
\includegraphics[width=0.9\linewidth]{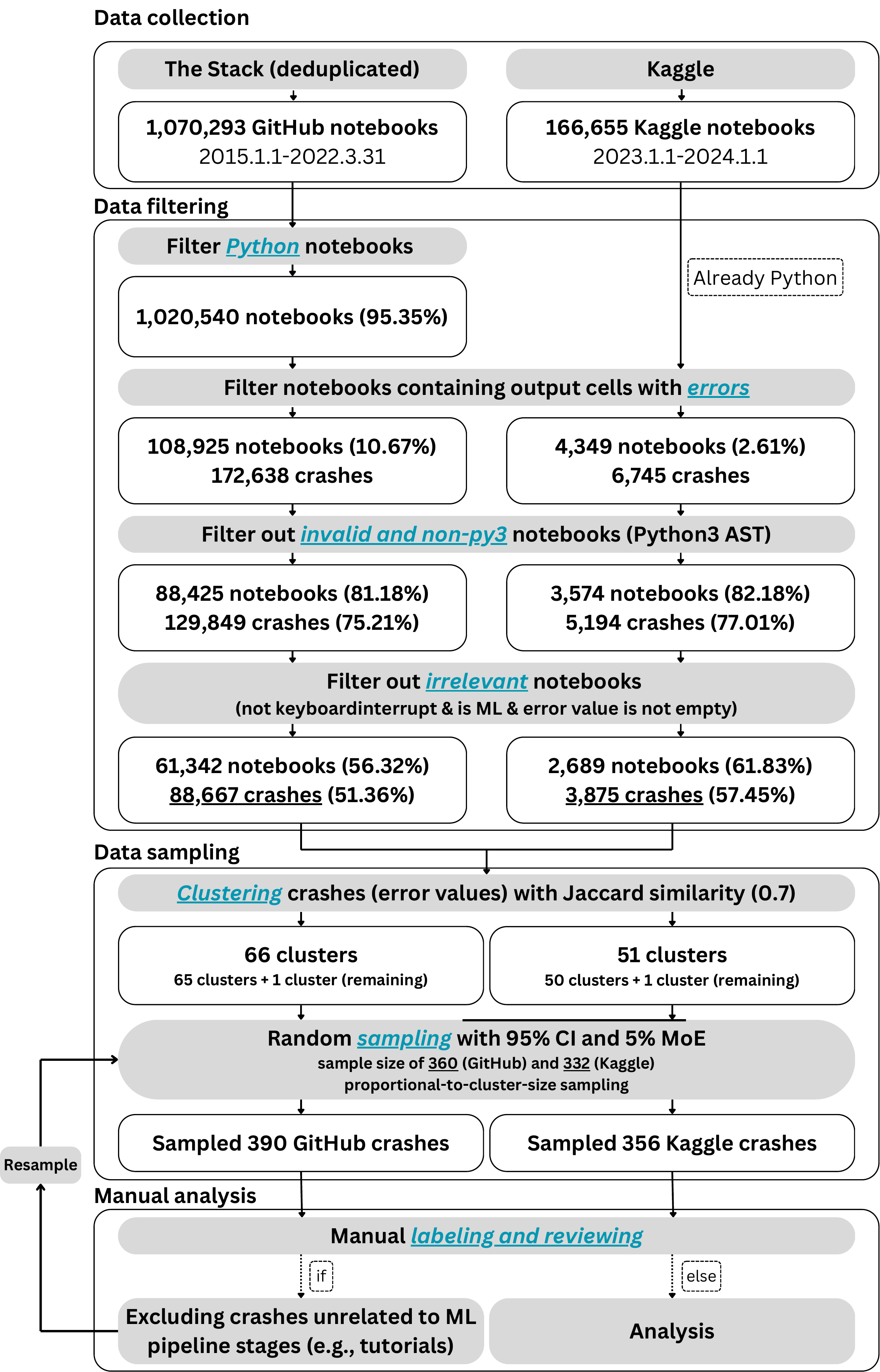}
\caption{Overview of the data collection, filtering, sampling, and manual analysis for Kaggle and GitHub notebooks.}
\label{fig: dataset}
\end{figure}


%% file: Content/Subcontents/tb_top_libs.tex

\begin{table}[b]
    \scriptsize
    \centering
    \caption{Top 24 popular ML libraries identified from 166,655 Kaggle notebooks with their usage frequencies, grouped by their main functionalities. DL libraries are marked in bold.}
    \begin{tabular}{m{2.1cm}|m{5.9cm}}
        \hline
        \thead{Theme}  & \thead{ML libraries} \\ \hline
        \makecell{Data manipulation\\ \& data analysis} 
        & 
        \makecell{\pandas(82.2\%),\numpy(81.4\%),\scipy(8.5\%),\\ \statsmodels(3.3\%)} \\
        \hline
        \makecell{Data visualization} 
        & 
        \makecell{\matplotlib(66.0\%),\seaborn(43.3\%),\plotly(8.4\%), \\ \wordcloud(2.1\%), \missingno(1.6\%) }\\
        \hline
        \makecell{ML frameworks\\ (build/train \\ ML models)} 
        &
        \makecell{\sklearn(52.8\%),\tensorflow(16.4\%),\keras(7.6\%),\\ \torch(10.0\%),\xgboost(8.6\%),\lightgbm(4.5\%), \\ \catboost(3.3\%),\imblearn(2.9\%) }\\
        \hline
        \makecell{Computer vision} 
        &
        \cvtwo(7.7\%),\torchvision(4.3\%),\skimage(1.5\%)\\
        \hline
        \makecell{NLP} 
        &
        \makecell{\nltk(4.3\%),\transformers(3.7\%)}\\
        \hline
        \makecell{Optimization \& \\ model tuning} 
        &
        \makecell{\optuna(1.5\%)} \\
        \hline
        \makecell{Dataset handling} 
        &
        \makecell{\datasets(1.4\%)} \\
        \hline
    \end{tabular}
    \label{tb: top_ML_libs}
\end{table}

%% file: Content/Subcontents/tb_jaccard_sim.tex
\begin{table}[t]
    \scriptsize
    \centering
    \caption{Jaccard similarity result (JS) of two error values.}
    \begin{tabular}{m{3.5cm}|m{3.5cm}|m{0.5cm}}
        \hline
        \thead{Error value (original)} & \thead{Error value to compare} & \thead{JS}\\
        \hline
        Input 0 of layer "sequential\_1" is incompatible with the layer: expected shape=(None, 32), found shape=(None, 38) &
        input of layer is incompatible with the expected shape none found &
        \multirow{5}{*}{0.79}\\\cline{1-2}
        Input 0 of layer sequential\_1 is incompatible with the layer: expected ndim=3, found ndim=2. Full shape received: (None, None) & 
        input of layer is incompatible with the expected ndim found full shape received none & \\
        \hline
    \end{tabular}
    \label{tb: ja_sim}
\end{table}

%% file: Content/Subcontents/tb_cluster_exp.tex
\begin{table}[b]
    \scriptsize
    \centering
    \caption{Different clusters within \texttt{Value Error} exception.}
    \begin{tabular}{m{0.9cm}|m{5.2cm}|m{1.5cm}}
        \hline
        \thead{Cluster} & \thead{Example error value} & \thead{Crash type}\\\hline
        2255 & operands could not be broadcast together with shapes (25,200) (25,50) & \makecell[c]{Unsupported\\broadcast}\\
        \hline
        559 & Classification metrics can't handle a mix of binary and continuous targets & \makecell[c]{Data value\\violation}\\
        \hline
        720 & Input 0 of layer "sequential" is incompatible with the layer: expected shape=(None, 4), found shape=(None, 128) & \multirow{5}{*}{\makecell[c]{Tensor shape\\mismatch}}\\
        \cline{0-1}
        6862 & cannot reshape array of size 0 into shape (2,2) & \\
        \cline{0-1}
        6473 & shapes (25,) and (26,1) not aligned: 25 (dim 0) != 26 (dim 0) & \\
        \hline
    \end{tabular}
    \label{tb: cluster_exp}
\end{table}

%% file: Content/Subcontents/tb_dimensions.tex
\begin{table*}[t]
    \scriptsize
    \centering
    \begin{minipage}{\textwidth}
    \caption{Classification dimensions of crashes in ML notebooks.}
    \begin{tabular}{m{0.5cm}|m{1.6cm}|m{4.8cm}|m{9.4cm}}
        \hline
        \thead{RQ} & \thead{Dimension} & \thead{Category} & \thead{Description}\\
        
        \hline
        \hline
        
        \multirow{35}{*}{RQ1} & \multirow{20}{*}{\catone} & 
        \makecell[l]{\texttt{Variable Not Found} (VNF)} &
        A referenced variable is not defined or accessible. A refined type of \texttt{Name Error}.\\
        \cline{3-4}
        & &
        \makecell[l]{\texttt{Invalid Argument} (IARG)} &
        A function receives an argument with an incorrect type, format, or value, or has missing or misordered arguments, violating API constraints.\\
        \cline{3-4}
        & &
        \makecell[l]{\texttt{IO Error} (IO)} &
        A crash occurs when reading from or writing to a file,~\eg~file/path not found, no permission to the file, invalid file type, or unable to parse or decode the file.\\
        \cline{3-4}
        & &
        \makecell[l]{\texttt{Module Not Found} (MODULE)} &
        A required library or module is missing or not installed. A refined type of \texttt{Name Error}.\\
        \cline{3-4}
        & &
        \makecell[l]{\texttt{Tensor Shape Mismatch} (TSHAPE)} &
        Tensor shapes are incompatible for the intended operation. A refined type of \texttt{Value Error}.\\
        \cline{3-4}
        & &
        \makecell[l]{\texttt{Data Value Violation} (DVIOL)} &
        Dataset contains unexpected or invalid values that violate API constraints or user assumptions. A refined type of \texttt{Value Error}.\\
        \cline{3-4}
        & &
        \makecell[l]{\texttt{Out of Memory} (OOM)} &
        The available memory of GPU/CPU is insufficient to execute the operation.\\
        \cline{3-4}
        & &
        \makecell[l]{\texttt{Environment Error} (ENV)} &
        Issues related to system setup or import failures.\\
        \cline{3-4}
        & &
        \makecell[l]{\texttt{Request Error} (RERR)} &
        A failure occurs when making an external API or network request.\\
        \cline{3-4}
        & &
        \makecell[l]{\texttt{Unsupported Broadcast} (BCAST)} &
        Broadcasting fails due to incompatible array or tensor shapes. A refined type of \texttt{Value Error}.\\
        \cline{3-4}
        & &
        \makecell[l]{\texttt{Feature Name Mismatch} (FNAME)} &
        Feature names do not match expected names or formats in ML model structure or dataset. A refined type of \texttt{Value Error}.\\
        \cline{3-4}
        & &
        \makecell[l]{\texttt{Model Initialization Error} (INIT)} &
        Crash during model creation due to incorrect parameters or architecture.\\
        \cline{3-4}
        & &
        \makecell[l]{Python built-in exception types\\(\eg~\texttt{Attribute Error}(ATTR),\\ \texttt{Key Error}(KEY), \texttt{Value Error}(VALUE),\\\texttt{Type Error}(TYPE), \texttt{Name Error}(NAME),\\\texttt{Index Error}(INDEX), \\\texttt{Runtime Error}(RUNTIME), \\\texttt{System Error}, \texttt{Zero Division Error})} &
        These are exceptions that occur during the execution of Python code, typically caused by issues like attribute access errors, illegal dictionary keys, invalid values, type mismatches, undefined function/class names, out-of-bounds indexing, execution failures, system-related issues, division by zero, and so on. More details can be found in the \hrefcus{https://docs.python.org/3/library/exceptions}{Python documentation}.\\ \cline{2-4}
        & \multirow{12}{*}{\cattwo} &
         
        API misuse (API) &
        Incorrect usage of an API,~\eg~passing invalid arguments or calling functions improperly.\\\cline{3-4}
        & & 
        Notebook-specific issue (NB) &
        Errors that stem from the unique execution behavior of notebooks,~\eg~out-of-order execution, and unresolved errors in prior cells (previous cell error).\\\cline{3-4}
        & & 
        Data confusion (DATA) &
        Misinterpretation of data shapes, structure, or values, leading to unexpected errors.\\\cline{3-4}
        & &
        Implementation error (IMPL) &
        Mistakes in code logic,~\eg~incorrect variable usage, function calls, or algorithm design.\\\cline{3-4}
        & & 
        Library issue (LIB)&
        Errors originating from external libraries, including bugs or deprecated APIs.\\\cline{3-4}
        & & 
        Environment issue (ENV)&
        Issues due to incorrect environment setting,~\eg~missing/incompatible dependencies, file path errors, or external control.\\\cline{3-4}
        & & 
        ML model Mismatch (MODEL) &
        A mismatch between a saved ML model and the expected model definition in the code.\\\cline{3-4}
        & & 
        Insufficient resource (RSC)&
        Crashes due to resource constraints,~\eg~running out of memory or disk space.\\\cline{3-4}
        & & 
        Unknown (UNK) &
        Errors for which the root cause cannot be determined.\\
        
        \hline
        \hline
        
        \multirow{4}{*}{RQ2} & \multirow{2}{*}{\catthree} &        
        ML bug &
        Errors in code using ML libraries or related components.\\\cline{3-4}
        & & 
        Python bug &
        Errors from general Python code, unrelated to ML libraries.\\\cline{2-4}

        & \multirow{2}{*}{\catfive} &
        Specific ML libraries &
        The crash is caused by using one/more ML libraries (see~\autoref{tb: top_ML_libs}).\\\cline{3-4}
        & & 
        None &
        The crash is neither ML-related bug nor caused by directly using any ML libraries.\\
        \hline
        \hline
        
        \multirow{7}{*}{RQ3} & \multirow{7}{*}{\catfour} & 
        Environment setup (ENVS) &
        Errors related to environment setup,~\eg~missing modules or configuration issues.\\\cline{3-4}
        & & 
        Data preparation (DATAP) &
        Errors occurring during data preprocessing or preparation for analysis and training.\\\cline{3-4}
        & & 
        Data visualization (DATAV) &
        Errors encountered while visualizing data.\\\cline{3-4}
        & & 
        Model construction (MCONS) &
        Issues related to building, compiling, or visualizing ML models.\\\cline{3-4}
        & & 
        Model training (TRAIN) &
        Errors that arise during model training, including hyperparameter tuning and validation.\\\cline{3-4}
        & & 
        Evaluation/ prediction (EVAL) &
        Errors related to evaluating model performance or making predictions.\\
        
        \hline
    \end{tabular}
    \label{tb: dimensions}
    \end{minipage}
\end{table*}

%% file: Content/Subcontents/tb_stats_manual_label.tex
\begin{table}[t]
    \scriptsize
    \centering
    \caption{Changes in the number and percentage of labeled crashes after reviewing the final of sampled crashes (746 crashes after resampling).}
    \begin{tabular}{m{0.5cm}|m{1.7cm}|m{1cm}|m{1.25cm}}
        \hline
        \thead{RQ} & \thead{Dimension}  & \thead{Number} & \thead{Percentage}\\ \hline
        \multirow{2}{*}{RQ1} & \catone    & 31         & 4.16\%   \\ 
                             & \cattwo    & 72         & 9.65\%   \\ \hline
        \multirow{2}{*}{RQ2} & \catthree  & 24         & 3.22\%   \\ 
                             & \catfive   & 32         & 4.29\%   \\ \hline
        RQ3                  & \catfour   & 64         & 8.58\%   \\ \hline
    \end{tabular}
    \label{tb: stats_manual_label}
\end{table}




%% file: Content/results.tex
\section{Results}
\label{sect: results}

Next, we present the results to answer each research question. To investigate correlations between two distributions, we use the $\chi^2$ test to assess whether there is a significant difference, with a null hypothesis stating no difference. We adopt a 95\% confidence level, meaning that a $p$-value less than 0.05 indicates a significant difference. For distributions with small sample sizes ($<5$), we apply Barnard's exact test~\cite{barnad1947test}, using Bonferroni correction~\cite{bonferroni1936teoria} when the categories are non-binary. When specifying our results, we refer to GitHub as \GH~and Kaggle as \KG.

\subsection{RQ1: \rqone} \label{sect: result_rq1}

\input{Content/Subcontents/fig_results_rq1_exception_types_and_root_causes}

Crashes in ML notebooks stem from various types of exceptions.
We refine and enhance the raw exception types typically shown in the crash information and form crash types.
We then complement these classes by identifying their underlying root causes and discussing their relations.
The results are visualized in \autoref{fig: res_rqone} and \autoref{fig: res_rqone_co_occurrence}.

\paragraph*{\textit{Crash types}}
\autoref{fig: res_exception_types} highlights the most frequent crash types in both GitHub and Kaggle and crash types that are unique to ML.
The most common crash type is \texttt{Variable Not Found}, accounting for 17.8\% of cases (\GH: 18.5\%, \KG: 17.1\%), followed by \texttt{Invalid argument} at 9.8\% (\GH: 10.3\%, \KG: 9.3\%), \texttt{IO Error} at 9.7\% (\GH: 9.0\%, \KG: 10.4\%), \texttt{Module Not Found} at 8.0\% (\GH: 9.7\%, \KG: 6.2\%), and \texttt{Attribute Error} at 7.0\% (\GH: 7.4\%, \KG: 6.5\%). 

\important{Finding 1.1}{The most common crash types of ML notebook crashes are \texttt{Variable Not Found} (17.8\%), \texttt{Invalid argument} (9.8\%), and \texttt{IO Error} (9.7\%). }

Moreover, crash types \texttt{Tensor Shape Mismatch} (5.5\%; \GH: 4.6\%, \KG: 6.5\%), \texttt{Data Value Violation} (3.5\%; \GH: 3.9\%, \KG: 3.1\%), \texttt{Out of Memory} (2.1\%; \GH: 1.3\%, \KG: 3.1\%), \texttt{Unsupported Broadcast} (1.6\%; \GH: 1.3\%, \KG: 2.0\%), \texttt{Feature Name Mismatch} (1.2\%; \GH: 1.5\%, \KG: 0.8\%), and \texttt{Model Initialization Error} (0.7\%; \GH: 0.5\%, \KG: 0.8\%) are ML-specific crash types.

The $\chi^2$ test shows no significant difference in crash type distributions between GitHub and Kaggle ($p>0.05$).

\important{Finding 1.2}{ML-specific crashes types (14.6\%) include \texttt{Tensor Shape Mismatch}, \texttt{Data Value Violation}, \texttt{Out of Memory}, \texttt{Unsupported Broadcast}, \texttt{Feature Name Mismatch}, and \texttt{Model Initialization Error}.}

\paragraph*{\textit{Root causes}}
\autoref{fig: res_root_causes} visualizes the distribution of root causes for GitHub and Kaggle separately.
The most common ones are API misuse (20.9\% overall; \GH: 21.0\%, \KG: 20.8\%), notebook-specific issues (19.4\%; \GH: 19.2\%, \KG: 19.7\%), implementation errors (16.9\%; \GH: 17.4\%, \KG: 16.3\%), environment issues (16.8\%; \GH: 18.0\%, \KG: 15.5\%), and data confusion (16.4\%; \GH: 16.9\%, \KG: 15.7\%).

\important{Finding 1.3}{The main root causes of ML notebook crashes include API misuse~(20.9\%), notebook-specific issues (19.4\%), implementation errors (16.9\%), environment problems (16.8\%), and data confusion (16.4\%).}

Notebook-specific issues are a major cause of crashes, ranking second on both GitHub and Kaggle. The majority of these crashes result from out-of-order execution (\GH: 60.0\%, \KG: 74.3\%), where cells are executed in an incorrect sequence. This occurs when users run notebook cells without considering dependencies. Another common cause is previous cell error (\GH: 40.0\%, \KG: 25.7\%), where users encounter crashes in earlier cells but continue executing dependent cells without resolving the issues, resulting in additional crashes.

The root cause distribution does not differ significantly between GitHub and Kaggle ($p>0.05$).

\important{Finding 1.4}{Notebook-specific issues are the second major cause of crashes in ML notebooks, with out-of-order execution being the most frequent (\GH: 60.0\%, \KG: 74.3\%).}

\paragraph*{\textit{Root causes of crash types}}
We find no statistical difference in root causes between GitHub and Kaggle, hence we merge them and plot crash type and root cause co-occurrences in \autoref{fig: res_rqone_co_occurrence}.

We observe that API misuse, implementation errors, and data confusion are the root causes of a wide range of crashes in ML notebooks. Notably, API misuse alone is responsible for at least 25\% of crashes observed along multiple crash types, including \texttt{Invalid Argument} (66\%), \texttt{Attribute Error} (44\%), \texttt{Type Error} (42\%), \texttt{Value Error} (35\%), \texttt{Data Value Violation} (31\%), and \texttt{Tensor Shape Mismatch} (29\%). Meanwhile, data confusion contributes to over 30\% of several crash types, such as \texttt{Key Error} (50\%), \texttt{Data Value Violation} (50\%), \texttt{Unsupported Broadcast} (50\%), \texttt{Index Error} (45\%), \texttt{Value Error} (42\%), \texttt{Tensor Shape Mismatch} (34\%), and \texttt{Feature Name Mismatch} (33\%). These findings indicate that debugging is likely to be difficult in such cases. Developers need to be attentive to the proper use of APIs and ensure that their data inputs are correctly formatted.

\important{Finding 1.5}{API misuse, implementation errors, and data confusion are main causes of a wide range of crash types.}

Certain crash types are dominantly linked to one specific root cause. For instance, \texttt{Out of Memory} is always raised due to insufficient resources. \texttt{IO Error}, \texttt{Environment Error}, and \texttt{Request Error} are related to environment problems. \texttt{Invalid Argument} is mainly due to API misuse and data confusion. Name-related errors such as \texttt{Variable Not Found}, \texttt{Name Error}, and \texttt{Module Not Found} are mostly connected to notebook-specific issues. This highlights how notebook semantics can introduce unique challenges that manifest in particular crash types. 

\important{Finding 1.6}{Name-related errors, in particular \texttt{Variable Not Found}, are mainly due to notebook-specific issues.}

\paragraph*{\textit{Evaluating root cause accuracy}}
To assess the accuracy of our root cause analysis, we randomly selected 30 crashes in Kaggle notebooks and attempted to reproduce and fix them. For each case, we download the input datasets (if available), execute a subset of code cells necessary to trigger the crash, localize and repair the faulty code to verify if the labeled root cause aligns with the actual fault. A crash is considered fixed if the previously crashing cell executes successfully while preserving the original intent. 

We successfully reproduced and fixed 20 out of the 30 crashes. Of the 10 unsuccessful cases, 8 lack input datasets, and the remaining two no longer crash. From the 20 cases, 18 ($90\%$) match the root cause identified by this study, which suggests that our approach is reliable. For the two misclassifications, one involves an index error due to out-of-order execution mislabeled as API misuse. The other is a tensor shape mismatch caused by faulty logic (\ie~implementation error) in earlier cells, incorrectly labeled as data confusion. The main challenge in labeling errors and root causes without executing the notebooks lies in mentally tracking the internal kernel state. This can lead to misunderstandings by labelers, especially when the code logic is complex.

\input{Content/Subcontents/fig_results_rq1_exception_type_and_root_causes_co_occurrence}

\subsection{RQ2: \rqtwo} \label{sect: result_rq2}

Crashes in GitHub and Kaggle notebooks can stem from either code that uses ML libraries or general Python code. We analyze the distribution of ML/Python bugs to understand ML-specific crashes. Additionally, we examine the root causes of both ML and Python bugs, as shown in~\autoref{fig: res_rqtwo_co_occurrence}, to gain deeper insights into their distinctions and similarities.
Furthermore, we examine the frequency of crashes directly linked to the use of ML libraries (see~\autoref{fig: res_rqone_library}) to identify which libraries present the greatest challenges.

\paragraph*{\textit{ML bugs vs. Python bugs}}
ML bugs account for 67.7\% of crashes overall in ML notebooks (GH: 62.8\%, KG: 73.0\%). The $\chi^2$ test reveals a significant difference in the distribution of ML and Python bugs between the two platforms ($p<0.01$).
These findings suggest that a substantial proportion of crashes on both platforms stems from code that interacts with ML libraries, with the impact being particularly pronounced in Kaggle notebooks.

\important{Finding 2.1}{On both platforms, we find significantly more ML bugs than Python bugs. Kaggle exhibits a more significant prevalence of ML-related crashes than GitHub.}

\paragraph*{\textit{Root causes of ML and Python bugs}}
As \autoref{fig: res_rqtwo_co_occurrence} illustrates, the primary causes of ML-related crashes are API misuse (25\%), notebook-specific issues (20\%), and data confusion (20\%). Environment setup problems (12\%) and implementation errors (11\%) are also big factors.
Furthermore, ML-specific bugs include unique root causes such as ML model mismatch (3\%), insufficient resources (3\%), and library issues (1\%).
In contrast, general Python crashes are mainly driven by implementation errors (29\%) and environment issues (28\%), followed by notebook-specific issues (18\%) and API misuse (13\%). The differences in root causes between ML and general Python bugs stem from the inherent complexity of working with ML models and data, as well as the common use of notebook semantics when performing ML tasks. While the saved states in notebooks improve usability, they can also increase the likelihood of crashes if not properly managed.

\important{Finding 2.2}{The primary causes of ML-related crashes are API misuse (25\%), notebook-specific issues (20\%), and data confusion (20\%), while general Python crashes are mostly caused by implementation errors (29\%) and environment problems (28\%).}

\input{Content/Subcontents/fig_results_rq2_ml_bugs}

The root cause distributions between ML and Python bugs differ significantly ($p<0.001$). 
Notably, the higher percentage of API misuse in ML-related crashes (25\%) compared to general Python crashes (12\%) underscores the challenges developers face in correctly using ML libraries.

We find no statistically significant differences between the platforms in the distribution of root causes for either ML bugs or Python bugs ($p>0.05$ for both).

\important{Finding 2.3}{API misuse is more prevalent in ML-related crashes (25\%) than in general Python crashes (12\%).}

\input{Content/Subcontents/fig_results_rq1_library}

\paragraph*{\textit{ML libraries}}
\autoref{fig: res_rqone_library} ranks the top 10 most commonly used ML libraries by their \textit{challenge factor} ($CF$), which measures how likely using a ML library is to cause crashes compared to how often it is used. The \textit{Challenge factor} is computed as $CF = Lib_{\text{crashes}}/Lib_{\text{usage}}$, where 1) $Lib_{\text{crashes}} = N_{\text{Lib\_crash}}/N_{\text{crashes}}$ is the proportion of crashes caused by using an ML library. 2) $Lib_{\text{usage}} = N_{\text{Lib\_use}}/N_{\text{notebooks}}$ represents how frequently the library appears in ML notebooks, counting each notebook only once per library. A higher $CF$ indicates a library is more crash-prone relative to its usage. In this ranking, we merge \texttt{tensorflow}~and~\texttt{keras} because \texttt{keras} has been tightly integrated into \texttt{tensorflow} since TensorFlow 2.0. Combining them provides a more accurate representation of their impact compared to other ML libraries.
ML libraries~\texttt{tensorflow/keras}, and~\texttt{torch}~exhibit the highest challenge factors (over 5), followed by~\texttt{sklearn} (2.8),~\texttt{pandas} (2.5), and~\texttt{torchvision} (1.3).

\important{Finding 2.4}{\texttt{tensorflow/keras} and~\texttt{torch}~are the top ML libraries most challenging to use correctly.}

\subsection{RQ3: \rqthree} \label{sect: result_rq3}
The ML pipeline outlines the key stages of ML programs. We analyze the frequency of crashes occurring in these stages during ML development in notebooks and determine how often crashes in a certain stage involve ML bugs. The results are shown in \autoref{fig: res_rqthree}.

\paragraph*{\textit{ML pipeline}} \label{para: result_rq3_pp}
\autoref{fig: res_mlpp1} shows that most crashes occur during the data preparation (33.0\% overall; \GH: 37.7\%, \KG: 27.8\%) stage, model training (19.4\%; \GH: 17.4\%, \KG: 21.6\%), evaluation/prediction (18.4\%; \GH: 14.6\%, \KG: 22.5\%), and data visualization (13.5\%; \GH: 14.1\%, \KG: 12.9\%) stages. Crashes during data preparation are likely due to the complexities of cleaning and processing datasets. Model training and evaluation crashes may result from incorrect settings, misaligned inputs, or resource limitations. Data visualization errors are likely caused by mismatched data types or improper library usage.

\important{Finding 3.1}{The majority of crashes occurs during data preparation (33.0\%), model training (19.4\%), and evaluation/prediction (18.4\%) stages in the ML pipeline.}

\input{Content/Subcontents/fig_results_rq3_pipeline_vs_ml_bug}

We then examine the differences between GitHub and Kaggle in the ML pipeline stages.
A $\chi^2$ test reveals a significant difference ($p<0.01$).
To understand which stages contribute more to this difference, we calculate the adjusted residuals~\cite{sharpe2015residual}, comparing observed crashes in each stage with the expected number of crashes under the null hypothesis (shown in \autoref{fig: res_mlpp_residuals}).
A larger absolute residual indicates a greater contribution.
The results suggest that GitHub contains more crashes in the environment setup and data preparation, likely due to the variability in user environments and dependencies. Kaggle, with pre-configured environment, faces more crashes during evaluation/prediction, possibly due to model or input setting issues, or resource limitations.

\important{Finding 3.2}{GitHub and Kaggle crashes are distributed differently across ML pipeline stages. GitHub has more during environment setup and data preparation, whereas Kaggle sees more during the evaluation/prediction stage.}

\paragraph*{\textit{ML/Python bugs per ML pipeline stage}}
The breakdown of ML and Python bugs in each pipeline stage reveals that ML bugs dominate critical stages, with over 80\% of crashes during model construction and training, and over 70\% during evaluation/prediction for both platforms.

\important{Finding 3.3}{ML-related crashes are dominant (over 70\%) in model construction, training, and evaluation/prediction stages of the ML pipeline.}

We apply Barnard's exact test to assess significant differences of ML/Python bug distributions between GitHub and Kaggle for each ML pipeline stage. 
The results show significant differences in the data preparation ($p<0.01$) stage.
In Kaggle, 70\% of crashes during data preparation are ML-related, whereas in GitHub, ML and Python bugs occur at similar rates. 

\important{Finding 3.4}{Kaggle has 1.4 times more ML-related crashes in the data preparation stage than GitHub.}

\paragraph*{\textit{Root causes of ML pipeline stages}}
We further analyze root cause distribution across ML pipeline stages (see~\autoref{fig: res_mlpp_rc}). Most stages display a range of root causes, including API misuse, notebook-specific issues, implementation errors, environment issues, and data confusion. Notably, in the environment setup stage, 84\% of crashes stem specifically from environment-related issues.

\input{Content/Subcontents/fig_results_rq3_pipeline_vs_rc}

%% file: Content/Subcontents/fig_results_rq1_exception_types_and_root_causes.tex
\begin{figure}[t]
    \centering
    \begin{subfigure}[b]{0.95\columnwidth}
        \centering
        \includegraphics[width=\textwidth]{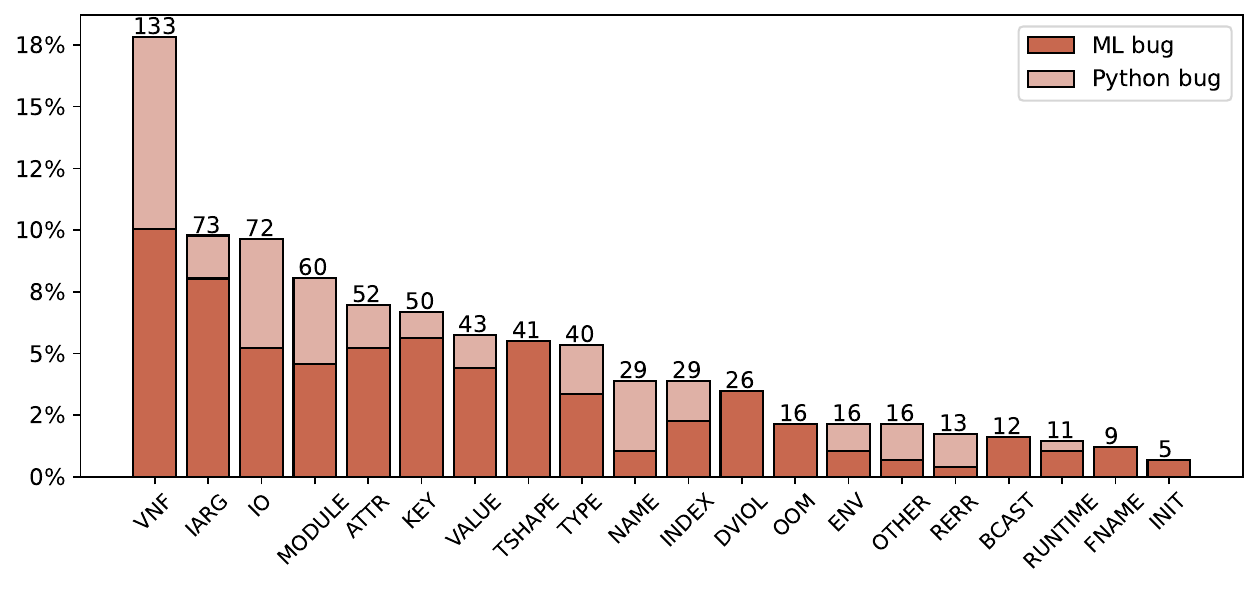} 
        \caption{Crash types (see \autoref{tb: dimensions} for the full names)}
        \label{fig: res_exception_types}
    \end{subfigure}
    \begin{subfigure}[b]{0.95\columnwidth}
        \centering
        \includegraphics[width=\textwidth]{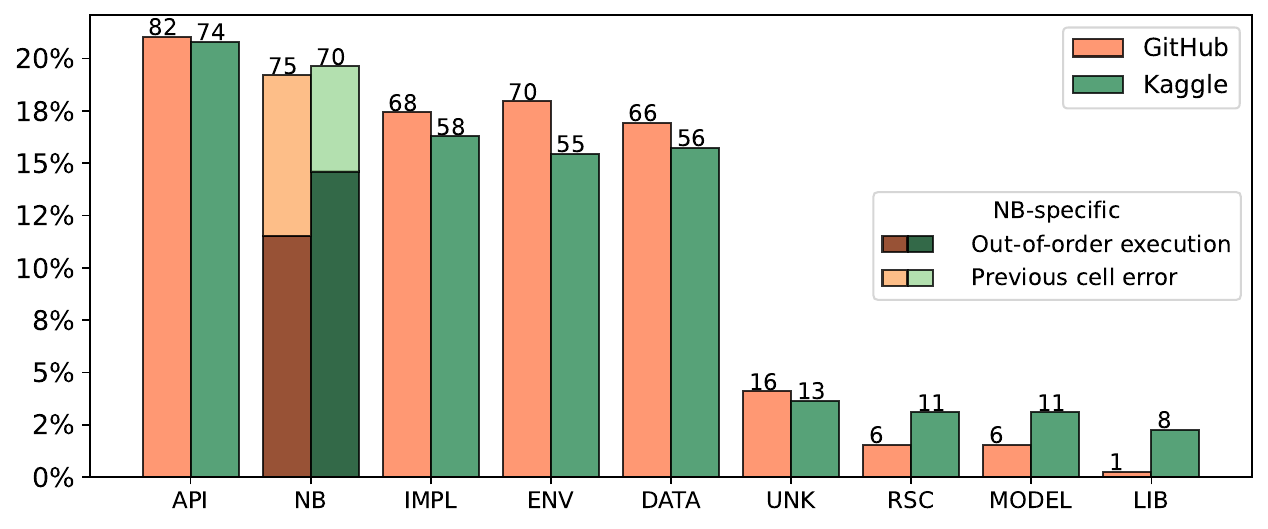} 
        \caption{Root causes (see \autoref{tb: dimensions} for the full names)}
        \label{fig: res_root_causes}
    \end{subfigure}
    \caption{Overview of the exception types and root causes of crashes in Jupyter notebooks from GitHub and Kaggle.}
    \label{fig: res_rqone}
\end{figure}

%% file: Content/Subcontents/fig_results_rq1_exception_type_and_root_causes_co_occurrence.tex
\begin{figure}[t]
    \centering
    \includegraphics[width=0.9\columnwidth]{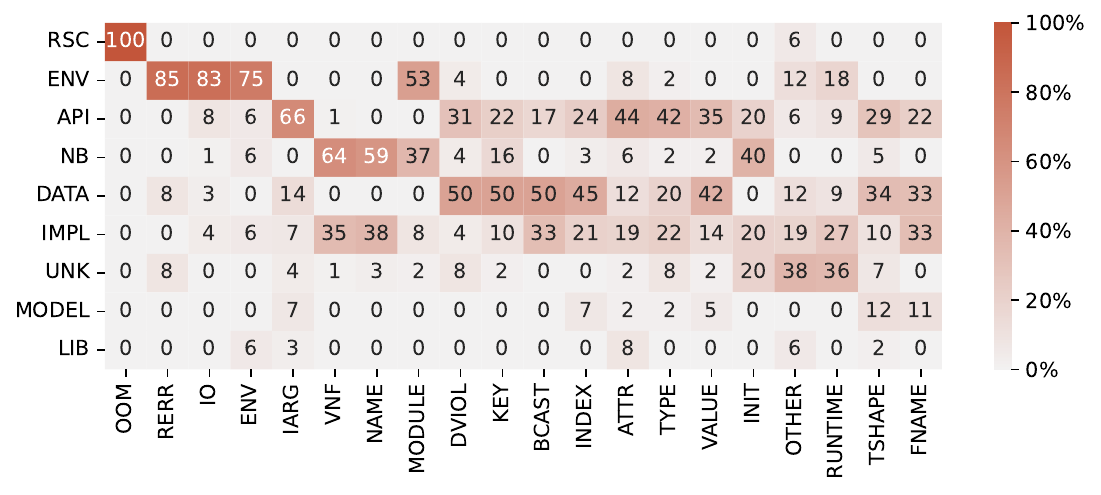} 
    \caption{Overview of the co-occurrence of crash types and root causes (see \autoref{tb: dimensions} for full names).}
    \label{fig: res_rqone_co_occurrence}
\end{figure}

%% file: Content/Subcontents/fig_results_rq2_ml_bugs.tex
\begin{figure}[b]
    \centering
    \includegraphics[width=0.95\columnwidth]{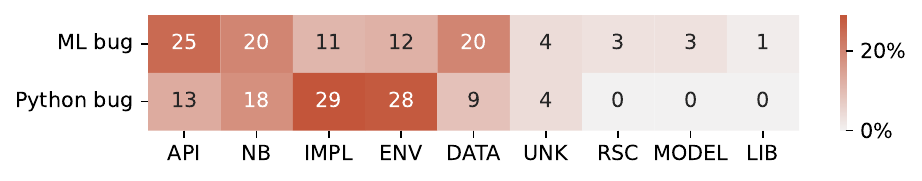} 
    \caption{Overview of the overall co-occurrence of ML/Python bugs and root causes (see \autoref{tb: dimensions} for full names).}
    \label{fig: res_rqtwo_co_occurrence}
\end{figure}

%% file: Content/Subcontents/fig_results_rq1_library.tex
\begin{figure}[t]
    \centering
    \includegraphics[width=1\columnwidth]{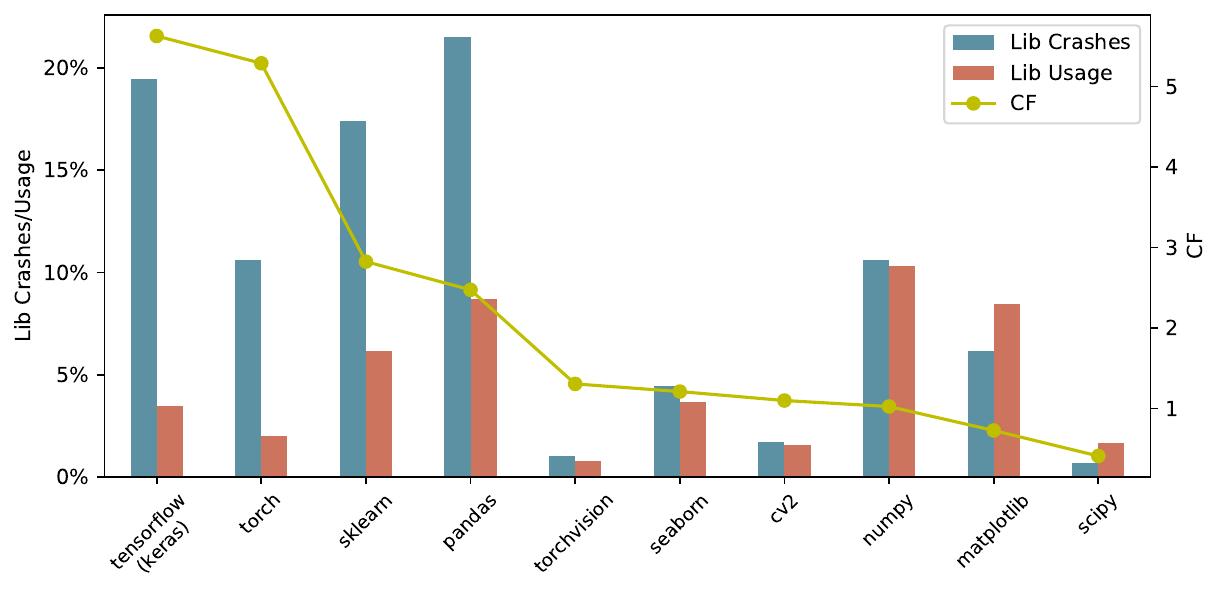}
    \caption{The challenge factors ($CF$) of the top 10 ML libraries, indicating the ratio between a library's involvement in crashes and its usage. See \autoref{tb: top_ML_libs} for library descriptions.}
    \label{fig: res_rqone_library}
\end{figure}

%% file: Content/Subcontents/fig_results_rq3_pipeline_vs_ml_bug.tex
\begin{figure}
    \centering
    \begin{subfigure}{0.64\linewidth}
        \centering
        \includegraphics[width=\linewidth]{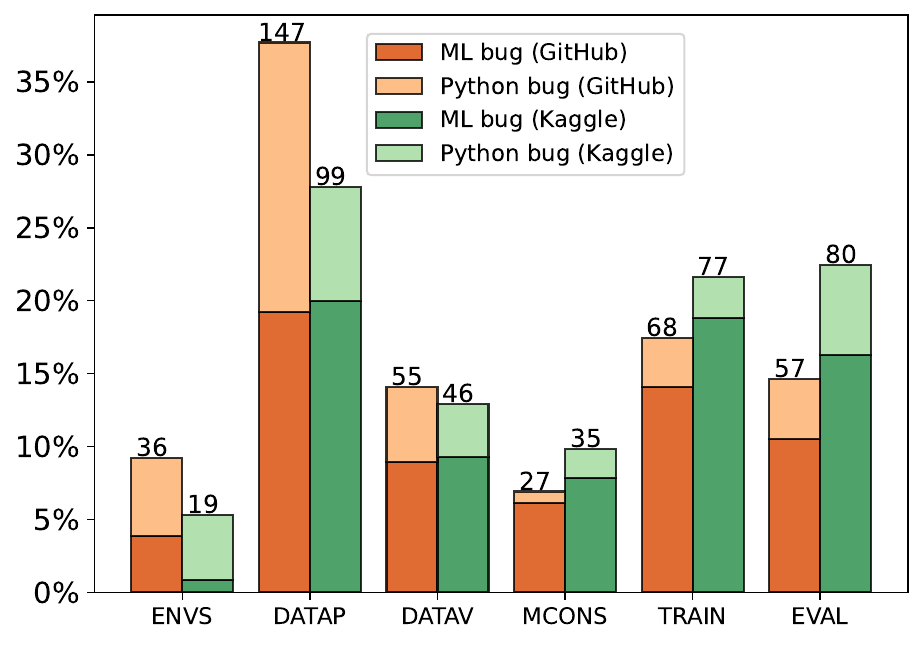} 
        \caption{ML pipeline stages}
        \label{fig: res_mlpp1}
    \end{subfigure}
    \begin{subfigure}{0.35\linewidth}
        \centering
        \includegraphics[width=\linewidth]{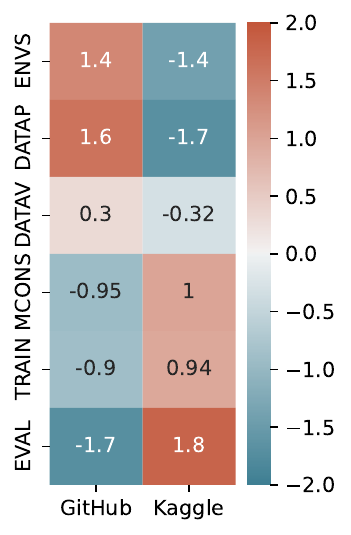} 
        \caption{Adjusted residuals}
        \label{fig: res_mlpp_residuals}
    \end{subfigure}
    \caption{Overview of crashes across the ML pipeline (see \autoref{tb: dimensions} for full names) with the distribution of ML/Python bugs per stage and the adjusted residuals per stage.}
    \label{fig: res_rqthree}
\end{figure}

%% file: Content/Subcontents/fig_results_rq3_pipeline_vs_rc.tex
\begin{figure}[t]
    \centering
    \includegraphics[width=.9\columnwidth]{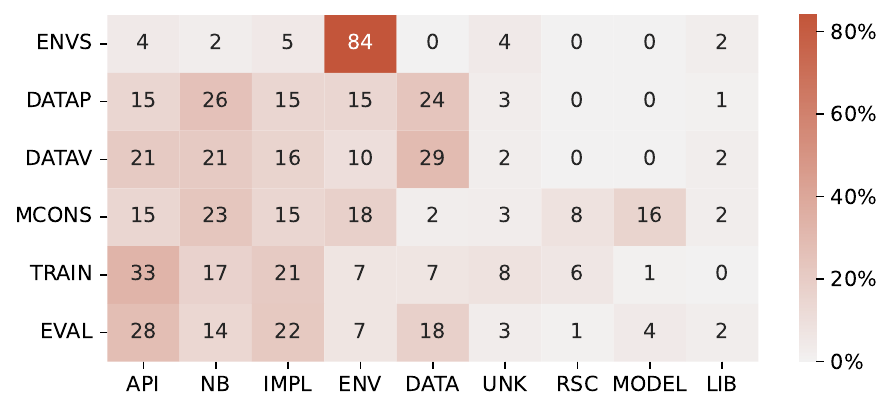}
    \caption{Overview of the overall co-occurrence of ML pipeline stages and root causes (see \autoref{tb: dimensions} for full names).}
    \label{fig: res_mlpp_rc}
\end{figure}

%% file: Content/discussion.tex
\section{Discussion}
\label{sect: discussion}

This section discusses the key implications of our findings and the differences between GitHub and Kaggle platforms. We then compare our findings with existing studies.

\subsection{Practical implications}

\paragraph*{\textit{Notebook-specific issues}}
As discussed in RQ1 (\autoref{sect: result_rq1}), notebook-specific issues contribute to a significant portion (19.4\%) of crashes, particularly those related to name resolution (e.g., \texttt{Variable Not Found}, \texttt{Name Error}, \texttt{Module Not Found}), \texttt{Model Initialization Error}, and key errors. These results highlight how the flexible execution model of notebooks can introduce instability, increasing the likelihood of crashes.

Unlike conventional software engineering workflows, where execution order is typically linear and well-defined, notebooks enable an interactive, fragmented development process. This flexibility, while beneficial for experimentation, also disrupts state management, leading to errors caused by out-of-order execution and implicit dependencies. For instance, if a variable \textit{x} is defined in one cell but a later cell that references \textit{x} is executed first, a \texttt{Variable Not Found} occurs due to out-of-order execution since \textit{x} has not been initialized yet. 
These issues underscore the unique challenges of debugging in notebooks, where errors may not be immediately apparent and can propagate unpredictably.

This finding underscores notebook development challenges, particularly in managing hidden state and execution order dependencies. Tools like NBSAFETY~\cite{Macke2021lineage} attempt to mitigate these risks by providing a custom Jupyter kernel that detects unsafe execution sequences. However, our results suggest that the problem remains prevalent, reinforcing the need for improved debugging support tailored to the dynamic execution model of notebooks.

\important{Takeaway 1}{The flexibility of notebooks disrupts state management, leading to unusual execution-order errors, which highlight the need for better debugging tools to support ML developers.}

\paragraph*{\textit{Data confusion}}
Another common root cause of crashes is data confusion (16.4\%), indicating practical challenges in handling heterogeneous data types.
There are many crash types, such as \texttt{Key Error}, \texttt{Data Value Violation}, \texttt{Unsupported Broadcast}, and \texttt{Index Error}, which point to the same root cause of data confusion. This potentially complicates fault localization process. Data confusion also accounts for a high percentage of ML-related crashes (20\%) (see \autoref{sect: result_rq2}), emphasizing the impact of improper data handling, such as incorrect transformations for ML APIs. Specifically, ML development often requires precise data formatting, and errors like shape mismatches or faulty transformations can be hard to allocate. This underscores the need for better data validation tools to prevent such issues.

\important{Takeaway 2}{Data confusion in ML notebooks leads to diverse crash types, making fault localization challenging.}

\paragraph*{\textit{ML bugs}}
When analyzing the results for RQ2 (\autoref{sect: result_rq2}), we find that crashes in ML notebooks are overall 2.1 times more frequently linked to ML bugs compared to Python bugs, reflecting key differences between developing ML and traditional software. Notably, a significantly higher proportion of ML-related crashes (25\%) are attributed to API misuse, compared to only 13\% for general Python crashes.



This issue stems from the heavy reliance on ML libraries, with complex usage guidelines~\cite{wei2024apimisuse, Galappaththi2024apimisuse} that often lead to misinterpretations of parameters and input-output expectations, causing crashes. Our analysis reveals that certain DL libraries, particularly \texttt{tensorflow/keras} and \texttt{torch}, pose greater challenges. In \texttt{tensorflow/keras}, tensor shape mismatches are a major cause of crashes, often due to implicit broadcasting, dynamic input shapes, or insufficient safeguards for detecting misaligned dimensions before execution. In contrast, \texttt{torch} is more prone to runtime errors, such as device compatibility issues (\eg~loading CUDA models on a CPU) and incorrect loss function configurations. Some of these errors may arise from unclear documentation, while others may stem from the flexibility of these libraries, where errors only surface at runtime, making static checks difficult. More informative error messages and warnings could significantly aid debugging.

These findings suggest that many crashes may stem not just from user mistakes but also from library design choices, including complex APIs, ambiguous error messages, and limited safeguards. This highlights the need for library developers to improve API usability, enhance documentation clarity, and implement better error-handling mechanisms. Additionally, identifying error-prone libraries can help practitioners prioritize debugging efforts, develop better tools, and choose libraries that align with their stability needs.

\important{Takeaway 3}{ML notebooks are prone to crashes due to API misuse, especially when using deep learning libraries \texttt{tensorflow/keras} and \texttt{torch}.}

\paragraph*{\textit{The data preparation stage}}
RQ3 (\autoref{sect: result_rq3}) demonstrates that ML notebook crashes occur mostly during the data preparation stage (33.0\%) of the ML pipeline.
This highlights the primary challenge developers face: preparing or preprocessing datasets to ensure compatibility with model requirements. Issues often arise from understanding the data itself or aligning it with API expectations. Additionally, the limited support available for developers during this stage implies that they may spend a disproportionate amount of time here. These insights suggest that improving tooling and resources for data preparation could significantly enhance developer productivity.

\important{Takeaway 4}{Data preparation is the most crash-prone stage in ML pipelines, highlighting the necessity for enhanced support to help developers navigate this critical phase.}

\paragraph*{\textit{The model training and evaluation/prediction stages}}
Crashes are more common in the model training (19.4\%) and evaluation/prediction (18.4\%) stages than in the model construction (8.3\%) stage.
These errors typically arise at runtime, when models are trained or evaluated with actual data inputs, rather than during the construction phase, when input shapes are often flexible. Libraries like \texttt{tensorflow/keras}, for instance, may not fully validate data compatibility until training begins, which can lead to errors surfacing only in these later stages. When issues occur within specific model layers, error messages can be ambiguous, making it difficult for developers to pinpoint and resolve problems. Consequently, developers may need to iteratively debug by re-training and re-evaluating multiple times, a tedious and time-intensive process.

Developers can avoid re-running all preceding cells as notebook semantics preserve previous cell executions, speeding up the iterative debugging process.
However, this also increases the risk of crashes from, for example, out-of-order execution, where dependencies may no longer align. 
These challenges highlight the importance of enabling early bug detection without requiring code execution for training or evaluating ML models, along with mechanisms to track cell dependencies and maintain the correct execution order.


\important{Takeaway 5}{Early bug detection through monitoring cell execution order in notebooks would catch issues before code execution, minimizing repetitive debugging cycles.}

\subsection{Comparison between GitHub and Kaggle}

\paragraph*{\textit{Crash rate}}
When filtering notebooks with crashes, we find that 10.67\% of GitHub Python notebooks contain error outputs, compared to 2.61\% on Kaggle. This difference likely reflects GitHub's broader use, such as version control and backup, where users upload more experimental and incomplete code, while Kaggle encourages sharing more complete solutions for ML competitions and tasks. 

\paragraph*{\textit{ML pipeline}}
Comparing ML notebook crashes between GitHub and Kaggle reveals their key differences. 
In the cases excluded from the initial sample, GitHub has a significantly higher proportion of crashes outside typical ML pipeline stages (25\% vs. 7\% on Kaggle), for example, tutorial notebooks, educational materials, or physics simulations. This suggests a more diverse use of notebooks where crashes often occur outside standard ML workflows.
Moreover, there is a notable difference in crash distributions across ML pipeline stages between the two platforms. For example, GitHub sees more crashes in data preparation, likely because it often uses “raw” datasets that require more complex preprocessing, whereas Kaggle provides standardized datasets with detailed descriptions. In contrast, Kaggle has more crashes during evaluation/prediction, possibly due to its submission process for competitions, which introduces additional model evaluation complexities.

\paragraph*{\textit{ML/Python bug}}
Kaggle shows a significantly higher prevalence of ML-related crashes than GitHub (73.0\% vs. 62.8\%).
One exception is during the environment setup stage of the ML pipeline, where
GitHub shows 2.6 times more ML-related crashes.
This is likely because Kaggle's pre-configured environment includes essential ML libraries, reducing setup-related errors. In contrast, GitHub, as a general-purpose code repository, preserves crashes from various environments.
During the data preparation stage, Kaggle notebooks face 1.4 times more ML-related crashes, suggesting that Kaggle users focus more on ML-specific tasks, while GitHub users encounter a broader range of coding challenges.
Furthermore, Kaggle provides a wealth of datasets alongside its notebooks, making it a valuable resource for reproducing experiments. However, existing studies~\cite{Pimentel19notebook, Chattopadhyay20notebook, Wang20notebook, Grotov22notebook, desantanaBugAnalysisJupyter2024} on notebooks have largely overlooked it as a data source.

\important{Takeaway 6}{Kaggle notebooks contain fewer error outputs than GitHub notebooks, but the crashes in Kaggle notebooks reflect challenges more specific to ML development, making them a more valuable resource in future research on the quality of ML notebooks.}

\subsection{Comparison with existing work}

To contextualize our results, we compare our findings with previous studies~\cite{zhangEmpiricalStudyProgram2020, desantanaBugAnalysisJupyter2024}. Zhang~\etal~\cite{zhangEmpiricalStudyProgram2020} identify failure types for DL \textit{jobs} similar to our crash types. They also report the distribution of failures across ML pipeline stages. De Santana~\etal~\cite{desantanaBugAnalysisJupyter2024} study bugs in Jupyter notebooks at a general level, which are similar to our target data format. 
To enable a fair comparison, we map our results to existing classifications of failure types~\cite{zhangEmpiricalStudyProgram2020} and bug types~\cite{desantanaBugAnalysisJupyter2024}. This is unnecessary for ML pipeline stages as our categorization includes the categories used in the other study~\cite{zhangEmpiricalStudyProgram2020}.

For the study by Zhang~\etal~\cite{zhangEmpiricalStudyProgram2020}, we retain only DL notebooks to match their domain. This filtering process follows a similar approach to identifying ML notebooks, as described in~\autoref{subsect: data_filtering}, where we compile a list of DL libraries as shown in~\autoref{tb: top_ML_libs}. Following the same methodology as ML library identification, two authors independently label the DL libraries, achieving a 100\% agreement rate. The overall mapped results are presented in~\autoref{tb: existing_work_compare}, while details on our mapping strategy and subcategory comparisons are available in our GitHub repository~\cite{wang2025code}.

\paragraph*{\textit{Bug type comparison with Zhang~\etal~\cite{zhangEmpiricalStudyProgram2020}}}
As~\autoref{tb: existing_work_compare_1} shows, execution environment errors are substantially lower in our DL dataset (compared to Zhang~\etal~\cite{zhangEmpiricalStudyProgram2020}). The authors attribute the high frequency of execution environment errors to differences between job execution environments and engineers' local environments. Our findings indirectly support this claim. Our target artifacts are DL notebooks typically executed either locally (GitHub) or in controlled environments with pre-installed dependencies (Kaggle). When developing DL notebooks in such stable environments, environment-related bugs are less common. Transitioning to other environments significantly increases the likelihood of encountering environment errors.

\important{Takeaway 7}{Crashes caused by execution environment are less frequent in ML notebooks but they can emerge when transitioning to execution platforms (\eg~cloud-based) used for deep learning. Practitioners should proactively test for such issues to ensure smooth deployment.}

After excluding the effect of execution environment errors,  our study still identifies a higher prevalence of DL-specific crashes. This discrepancy can be attributed to the different development stages being analyzed. Our results derive from examining cell outputs in notebooks primarily used for prototyping and early-stage DL development. In contrast, Zhang~\etal~mainly focus on later development stages when jobs (tested locally) are already being submitted to a DL platform. As such, our study should better represent the initial difficulties faced by DL developers.

\important{Takeaway 8}{As ML notebooks are used in early stage of development, deep learning-specific crashes are more prevalent in ML notebooks compared to deep learning jobs.}

\paragraph*{\textit{Bug type comparison with De Santana~\etal~\cite{desantanaBugAnalysisJupyter2024}}}
As shown in~\autoref{tb: existing_work_compare_2}, De Santana~\etal~\cite{desantanaBugAnalysisJupyter2024} find that environment and settings errors are considerably more common in Jupyter notebooks, as observed in GitHub commits, compared to our findings.
This discrepancy is likely due to differences in the nature of the analyzed notebooks. Specifically, their dataset primarily consists of notebooks from stable and mature GitHub projects, such as AWS, Azure, Fastai, and TensorFlow, whereas our dataset focuses on notebooks used for ML prototyping and experiments. To better understand this difference, we randomly sample 20 commits from their dataset to observe that the associated notebooks are predominantly demonstrations, documentation, or usage examples of specific libraries or software created by the developers of those libraries. The corresponding commits are mostly aimed at updating dependencies required by the projects rather than fixing implementation bugs, reflecting the stable and mature nature of these projects. As a result, it is expected that environment and settings issues are more prevalent in their dataset. 


We believe that our study better reflects the practical development challenges in notebook environments. We observe unique implementation errors, particularly those related to \textit{notebook-specific issues} like out-of-order execution and previous cell errors. While out-of-order execution has been linked to reproducibility concerns in prior research~\cite{Pimentel19notebook}, its role as a root cause of crashes has not been widely explored.
By analyzing crash information, code, and cell outputs, our study also distinguishes crashes caused by \textit{data confusion} from those resulting from API misuse. Error messages and outputs provide crucial insights into data shapes and structures, helping us pinpoint the origin of crashes more effectively. Therefore, we present these implementation issues separately in \autoref{tb: existing_work_compare_2} by marking them as “Implementation (other).”
 
\input{Content/Subcontents/tb_existing_work_compare}

\important{Takeaway 9}{The error patterns prevalent in Jupyter notebooks created by authors of ML libraries are different from ML notebooks created by the users of such libraries. }

\paragraph*{\textit{ML pipeline comparison with Zhang~\etal~\cite{zhangEmpiricalStudyProgram2020}}}
As~\autoref{tb: existing_work_compare_3} shows, Zhang~\etal~\cite{zhangEmpiricalStudyProgram2020} report considerably more environment setup failures but fewer training failures than our results. Their higher environment setup failure rate aligns with their larger proportion of execution environment bugs (\autoref{tb: existing_work_compare_1}). Fewer training errors may reflect that DL developers focus more on ensuring training correctness when developing locally before job submission.


As a novel finding, our analysis reveals that data visualization can also be a source of crashes in ML workflows because notebooks serve not only for coding but also for documentation and exploratory data analysis. The omission of this stage in prior work suggests that visualization-related crashes might be underrepresented. Identifying this category highlights the unique nature of notebook development and underscores the need for debugging support tailored to interactive workflows.

\important{Takeaway 10}{Since ML notebooks blend coding with exploration and documentation, notebook development has some unique types of errors uncommon in traditional ML workflows, particularly, in data visualization.}

%% file: Content/Subcontents/tb_existing_work_compare.tex
\begin{table}[t]
    \centering
    \begin{subtable}{1\linewidth}
        \scriptsize
        \centering
        \caption{Bug type classification comparison with Zhang~\etal~(S1~\cite{zhangEmpiricalStudyProgram2020}). The last column shows normalized percentages excluding execution environment (Env.). Our results are based on DL-only notebooks crashes.}
        \begin{tabular}{m{2.5cm}|m{1.3cm}|m{1.3cm}|m{1.8cm}} %
            \hline
            \thead{Dimensions\\S1~\cite{zhangEmpiricalStudyProgram2020}} 
            & \thead{Result\\S1~\cite{zhangEmpiricalStudyProgram2020}} & \thead{Result\\Ours (DL)}
            & \thead{Exclude Env.\\S1~\cite{zhangEmpiricalStudyProgram2020} vs. Ours}
            \\\hline
            Execution environment 
            & \makecell[c]{48.0\%} & \makecell[c]{19.9\%} & \makecell[c]{-} \\\hline
            Data                  
            & \makecell[c]{2.0\%}  & \makecell[c]{0.9\%} & \makecell[c]{3.8\% vs. 1.1\%} \\\hline
            DL specific           
            & \makecell[c]{13.5\%} & \makecell[c]{32.5\%} & \makecell[c]{26.0\% vs. 40.6\%} \\\hline
            General code error    
            & \makecell[c]{36.5\%} & \makecell[c]{46.7\%} & \makecell[c]{70.2\% vs. 58.3\%} \\\hline
        \end{tabular}
        \label{tb: existing_work_compare_1}
    \end{subtable}
    \begin{subtable}{1\linewidth}
        \scriptsize
        \centering
        \caption{Bug type classification comparison with De Santana~\etal~(S2~\cite{desantanaBugAnalysisJupyter2024}). We present our mapped results separately for Kaggle and GitHub for fair comparison.}
        \begin{tabular}{m{2.8cm}|m{1.45cm}|m{1.35cm}|m{1.35cm}} %
            \hline
            \thead{Dimensions\\S2~\cite{desantanaBugAnalysisJupyter2024}}
            & \thead{Result\\S2~\cite{desantanaBugAnalysisJupyter2024} (GH)} & \thead{Result\\Ours (KG)} & \thead{Result\\Ours (GH)} \\\hline
            Environment and settings
            &  \makecell[c]{35.6\%} & \makecell[c]{3.4\%} & \makecell[c]{6.9\%} \\\hline
            Processing              
            & \makecell[c]{1.9\%}  & \makecell[c]{3.1\%} & \makecell[c]{1.5\%}  \\\hline
            Implementation          
            & \makecell[c]{44.2\%} & \makecell[c]{26.4\%} & \makecell[c]{23.1\%} \\\hline
            \makecell[l]{Implementation (other)} 
            & \makecell[c]{-}  & \makecell[c]{67.1\% }& \makecell[c]{68.5\%} \\\hline
            Other (e.g., Kernel) 
            &  \makecell[c]{18.3\%} & \makecell[c]{-} & \makecell[c]{-}\\\hline
        \end{tabular}
        \label{tb: existing_work_compare_2}
    \end{subtable}
    \begin{subtable}{1\linewidth}
        \scriptsize
        \centering
        \caption{ML pipeline stage comparison with S1~\cite{zhangEmpiricalStudyProgram2020} (DL cases only). Our normalized percentages exclude data visualization (DV) and model construction (MC) stages absent in S1.}
        \begin{tabular}{m{2.4cm}|m{1.2cm}|m{2cm}|m{1.2cm}}
            \hline
            \thead{ML pipeline stages\\Our category} 
            & \thead{Result\\S1~\cite{zhangEmpiricalStudyProgram2020}} 
            & \thead{Exclude DV\&MC\\ Ours (DL)} 
            & \thead{Result\\Ours (DL)}\\\hline
            Environment setup       
            & \makecell[c]{30.8\%} & \makecell[c]{10.0\%}       
            & \makecell[c]{7.6\%}   \\\hline
            Data preparation        
            & \makecell[c]{30.0\%}  & \makecell[c]{22.7\%}       
            & \makecell[c]{17.3\%} \\\hline
            Data visualization      
            &\makecell[c]{-}
            &\makecell[c]{-}
            & \makecell[c]{7.6\%}   \\\hline
            Model construction      
            &\makecell[c]{-}&\makecell[c]{-}
            & \makecell[c]{16.4\%}  \\\hline
            Training                
            & \makecell[c]{15.0\%} & \makecell[c]{35.7\%}        
            & \makecell[c]{27.2\%}  \\\hline
            Evaluation/prediction   
            & \makecell[c]{24.3\%} & \makecell[c]{31.6\%}        
            & \makecell[c]{24.0\%}  \\\hline
        \end{tabular}
        \label{tb: existing_work_compare_3}
    \end{subtable}
    \caption{Comparison with existing studies~\cite{zhangEmpiricalStudyProgram2020, desantanaBugAnalysisJupyter2024}. An overview of these studies is provided in~\autoref{tb: relatedwork}.}
    \label{tb: existing_work_compare}
\end{table}

%% file: Content/validity.tex
\section{Threats to validity}
\label{sect: validity}

This section outlines potential threats to validity, addressing factors that may impact our findings.

\subsection{External validity}
\paragraph*{\textit{Data selection bias}}
As our dataset only includes publicly shared GitHub and Kaggle notebooks, it may exclude crashes that users fix before publishing, thus crashes encountered during development may be underrepresented. As a result, our findings reflect crashes that persist, which may not be the full spectrum of debugging challenges. Future work could address this by analyzing version histories or autosaved checkpoints to capture a more complete development process.
Furthermore, although we have filtered out tutorial notebooks, the ML notebooks in our dataset may still not fully represent those used as prototypes in an industrial setting. This introduces a potential threat to validity when generalizing crash patterns that could propagate from notebooks to production. Future work should address this by studying ML prototype notebooks developed within industry to better assess how our findings apply.

\paragraph*{\textit{Temporal limitations in datasets}}
The GitHub dataset includes notebooks from January 1, 2015, to March 31, 2022, while the Kaggle dataset spans from January 1, 2023, to January 1, 2024. Given the rapid evolution of coding practices and tools used over time, notebooks from earlier years may reflect outdated coding styles or libraries. To mitigate this, our filtering process removes notebooks using older Python versions, improving the generalizability of the dataset.



\paragraph*{\textit{KeyboardInterrupt exceptions}}
In our analysis, we exclude \texttt{KeyboardInterrupt} exceptions as these typically reflect user intervention rather than code issues.
However, these errors could indicate infinite loops or unintended resource-heavy operations.
By discarding these exceptions, we may overlook cases where such issues manifest.

\subsection{Internal validity}
\paragraph*{\textit{Clustering crashes}}
We cluster crashes with similar patterns based on their error values.
We enhance this by removing specific substrings, like key names and numeric values.
However, this results in an empty string in cases where error values only contain these specific substrings, which are consequently grouped into the same cluster.
Examples are \texttt{KeyError} and \texttt{SystemExit}, which are very different errors.
Although a theoretical threat, an investigation indicates that it only accounts for 0.38\% of the dataset.

\paragraph*{\textit{Sampling limitations}}
Our proportional stratified sampling ensures broad representation, but rare strata may still be underrepresented despite a minimum sample size of one per stratum. Increasing this minimum was impractical due to manual labeling constraints.  To assess potential gaps, we conduct an exploratory analysis by manually examining a maximum of 10 crashes from 10 randomly selected small clusters in GitHub and Kaggle datasets separately. This reveals no previously unknown crash patterns, suggesting that our approach captures key crash types and root causes, although some rare errors may still be missed. Furthermore, cases excluded during resampling reduce certain strata sizes, which may impact the original sampling proportions.

\paragraph*{\textit{Manual labeling}}
We manually label ML notebook crashes, which can be prone to errors.
To enhance the reliability of classification, we use a structured two-stage process.
In the labeling stage, three evaluators independently and iteratively categorize crashes, followed by discussions to synchronize findings, reduce bias, and promote consistency.
In the reviewing stage, a reviewer validates the labels.
Any disagreements are resolved by consensus.

\paragraph*{\textit{Reproducibility of crashes and root cause analysis}}
The error outputs we analyze represent real execution crashes, but they are not necessarily reproducible due to factors such as execution order, missing datasets, or environment replication~\cite{Pimentel19notebook}.
While reproducing crashes could further validate our findings, it is challenging, especially for GitHub notebooks. Kaggle notebooks, with their accessible datasets and pre-configured environments, may enable reproducibility. Nevertheless, we carried out a small-scale validation to assess the accuracy of our root cause analysis by trying to reproduce a few crashes in Kaggle, and the high accuracy suggests that our approach is reliable (see~\autoref{sect: result_rq1}).
In future work, we plan to extend crash reproducibility efforts in Kaggle notebooks to validate our results thoroughly and create a reproducible dataset for deeper analysis.

%% file: Content/conclusion.tex
\section{Conclusions and Future Work}
\label{sect: conclusion}

This paper provides the first large-scale empirical study focusing on software crashes in public ML notebooks collected from GitHub and Kaggle. Our analysis classifies (1) crash types, (2) root causes, (3) ML pipeline stages, and (4) involved ML libraries. (5) We also distinguish between ML-specific and general Python bugs. 

Our findings reveal that the (1) most common crashes in ML notebooks are \texttt{Variable Not Found}, \texttt{Invalid Argument}, and \texttt{IO Error}. We have also identified prevalent ML-specific crash types such as \texttt{Tensor
Shape Mismatch}, \texttt{Data Value Violation}, and \texttt{Out of Memory}. Moreover, (2) the primary root causes of crashes are API misuse and notebook-specific issues (specifically, out-of-order execution and previous cell error). (3) Most crashes (70.8\%) occur during data preparation, model training, and evaluation/prediction stages of the ML pipeline, while data visualization bugs are observed for the first time.
(4) We also identify that certain ML libraries are more challenging to use for ML notebook development, in fact, crashes mostly occur when using DL libraries \texttt{tensorflow/keras} and \texttt{torch}. Finally, (5) ML bugs account for more crashes than Python bugs, especially in Kaggle notebooks.
Future work will expand on these findings by constructing a reproducible dataset of ML crashes, and investigate the temporal evolution of crash patterns.

%% file: main.bbl
\begin{thebibliography}{10}
\providecommand{\url}[1]{#1}
\csname url@samestyle\endcsname
\providecommand{\newblock}{\relax}
\providecommand{\bibinfo}[2]{#2}
\providecommand{\BIBentrySTDinterwordspacing}{\spaceskip=0pt\relax}
\providecommand{\BIBentryALTinterwordstretchfactor}{4}
\providecommand{\BIBentryALTinterwordspacing}{\spaceskip=\fontdimen2\font plus
\BIBentryALTinterwordstretchfactor\fontdimen3\font minus \fontdimen4\font\relax}
\providecommand{\BIBforeignlanguage}[2]{{%
\expandafter\ifx\csname l@#1\endcsname\relax
\typeout{** WARNING: IEEEtran.bst: No hyphenation pattern has been}%
\typeout{** loaded for the language `#1'. Using the pattern for}%
\typeout{** the default language instead.}%
\else
\language=\csname l@#1\endcsname
\fi
#2}}
\providecommand{\BIBdecl}{\relax}
\BIBdecl

\bibitem{Wang20notebook}
J.~Wang, L.~Li, and A.~Zeller, ``Better code, better sharing: on the need of analyzing {Jupyter} notebooks,'' in \emph{Proceedings of the ACM/IEEE 42nd International Conference on Software Engineering: New Ideas and Emerging Results}, 2020.

\bibitem{Pimentel19notebook}
J.~F. Pimentel, L.~Murta, V.~Braganholo, and J.~Freire, ``A large-scale study about quality and reproducibility of {Jupyter} notebooks,'' in \emph{2019 IEEE/ACM 16th International Conference on Mining Software Repositories}, 2019.

\bibitem{Koenzen20notebook}
A.~P. Koenzen, N.~A. Ernst, and M.-A.~D. Storey, ``Code duplication and reuse in {Jupyter} notebooks,'' in \emph{2020 IEEE Symposium on Visual Languages and Human-Centric Computing}, 2020, p. 1–9.

\bibitem{jahani2023data}
H.~Jahani, R.~Jain, and D.~Ivanov, ``Data science and big data analytics: a systematic review of methodologies used in the supply chain and logistics research,'' \emph{Annals of Operations Research}, 2023.

\bibitem{sohail2023genetic}
A.~Sohail, ``Genetic algorithms in the fields of artificial intelligence and data sciences,'' \emph{Annals of Data Science}, vol.~10, no.~4, p. 1007–1018, 2021.

\bibitem{Grotov22notebook}
K.~Grotov, S.~Titov, V.~Sotnikov, Y.~Golubev, and T.~Bryksin, ``A large-scale comparison of {Python} code in {Jupyter} notebooks and scripts,'' in \emph{Proceedings of the 19th International Conference on Mining Software Repositories}, 2022, p. 353–364.

\bibitem{swc2024}
{Software Center, Sweden}. (2024) Software center. \url{https://www.software-center.se/partners/}. Accessed: 2025-02-27.

\bibitem{nasahecc2024}
{NASA high-end computing capability}. (23 Apr, 2025) Using {Jupyter} notebook for machine learning development on {NAS} systems. \url{https://www.nas.nasa.gov/hecc/support/kb/using-jupyter-notebook-for-machine-learning-development-on-nas-systems_576.html}. Accessed: 2025-05-15.

\bibitem{islamComprehensiveStudyDeep2019}
M.~J. Islam, G.~Nguyen, R.~Pan, and H.~Rajan, ``A comprehensive study on deep learning bug characteristics,'' in \emph{Proceedings of the 2019 27th ACM Joint Meeting on European Software Engineering Conference and Symposium on the Foundations of Software Engineering}, 2019, p. 510–520.

\bibitem{desantanaBugAnalysisJupyter2024}
T.~L. De~Santana, P.~A. D. M.~S. Neto, E.~S. De~Almeida, and I.~Ahmed, ``Bug analysis in {Jupyter} notebook projects: An empirical study,'' \emph{ACM Transactions on Software Engineering and Methodology}, vol.~33, no.~4, p. 1–34, 2024.

\bibitem{zhangEmpiricalStudyTensorFlow2018}
Y.~Zhang, Y.~Chen, S.-C. Cheung, Y.~Xiong, and L.~Zhang, ``An empirical study on {TensorFlow} program bugs,'' in \emph{Proceedings of the 27th ACM SIGSOFT International Symposium on Software Testing and Analysis}, 2018, p. 129–140.

\bibitem{humbatovaTaxonomyRealFaults2020}
N.~Humbatova, G.~Jahangirova, G.~Bavota, V.~Riccio, A.~Stocco, and P.~Tonella, ``Taxonomy of real faults in deep learning systems,'' in \emph{Proceedings of the ACM/IEEE 42nd International Conference on Software Engineering}, 2020.

\bibitem{zhangEmpiricalStudyProgram2020}
R.~Zhang, W.~Xiao, H.~Zhang, Y.~Liu, H.~Lin, and M.~Yang, ``An empirical study on program failures of deep learning jobs,'' in \emph{Proceedings of the ACM/IEEE 42nd International Conference on Software Engineering}, 2020, p. 1159–1170.

\bibitem{Morovati24bug}
M.~M. Morovati, A.~Nikanjam, F.~Tambon, F.~Khomh, and Z.~M. Jiang, ``Bug characterization in machine learning-based systems,'' \emph{Empirical Software Engineering}, vol.~29, no.~1, 2023.

\bibitem{wang2025code}
Y.~Wang, W.~Meijer, J.~A.~H. López, U.~Nilsson, and D.~Varro. (2024) Experimentation artifacts of paper "why do machine learning notebooks crash? an empirical study on public python jupyter notebooks". \url{https://github.com/PELAB-LiU/jupyter_nbs_empirical}.

\bibitem{wang2025data}
------. (2025) Dataset of paper "why do machine learning notebooks crash? an empirical study on public python jupyter notebooks". \url{https://doi.org/10.5281/zenodo.14070487}.

\bibitem{ahmedCharacterizingBugsPython2023}
S.~Ahmed, M.~Wardat, H.~Bagheri, B.~D. Cruz, and H.~Rajan, ``Characterizing bugs in {Python} and {R} data analytics programs,'' in \emph{arXiv:2211.15533}, 2023.

\bibitem{thung12bugml}
F.~Thung, S.~Wang, D.~Lo, and L.~Jiang, ``An empirical study of bugs in machine learning systems,'' in \emph{2012 IEEE 23rd International Symposium on Software Reliability Engineering}, 2012.

\bibitem{sunEmpiricalStudyReal2017}
X.~Sun, T.~Zhou, G.~Li, J.~Hu, H.~Yang, and B.~Li, ``An empirical study on real bugs for machine learning programs,'' in \emph{2017 24th Asia-Pacific Software Engineering Conference}, 2017.

\bibitem{jiaEmpiricalStudyBugs2020}
L.~Jia, H.~Zhong, X.~Wang, L.~Huang, and X.~Lu, \emph{An Empirical Study on Bugs Inside {TensorFlow}}.\hskip 1em plus 0.5em minus 0.4em\relax Springer International Publishing, 2020, p. 604–620.

\bibitem{chenUnderstandingDeepLearning2023}
J.~Chen, Y.~Liang, Q.~Shen, J.~Jiang, and S.~Li, ``Toward understanding deep learning framework bugs,'' \emph{ACM Transactions on Software Engineering and Methodology}, vol.~32, no.~6, p. 1–31, 2023.

\bibitem{avizienis2004faultdef}
A.~Avizienis, J.-C. Laprie, B.~Randell, and C.~Landwehr, ``Basic concepts and taxonomy of dependable and secure computing,'' \emph{IEEE Transactions on Dependable and Secure Computing}, vol.~1, no.~1, p. 11–33, 2004.

\bibitem{Kocetkov2022TheStack}
D.~Kocetkov, R.~Li, L.~B. Allal, J.~Li, C.~Mou, C.~M. Ferrandis, Y.~Jernite, M.~Mitchell, S.~Hughes, T.~Wolf, D.~Bahdanau, L.~von Werra, and H.~de~Vries, ``{The} {Stack}: 3 {TB} of permissively licensed source code,'' in \emph{arXiv:2211.15533}, 2022.

\bibitem{KGTorrent_Quaranta}
L.~Quaranta, F.~Calefato, and F.~Lanubile, ``{KGTorrent}: A dataset of {Python} {Jupyter} notebooks from {Kaggle},'' in \emph{2021 IEEE/ACM 18th International Conference on Mining Software Repositories}, 2021.

\bibitem{megan2022metakaggle}
M.~Risdal and T.~Bozsolik, ``{Meta} {Kaggle},'' 2022, https://www.kaggle.com/ds/9.

\bibitem{guesslang2024}
(2024) Guesslang. \url{https://github.com/yoeo/guesslang}. Accessed: 2024-06-13.

\bibitem{zhang2021study}
H.~Zhang, S.~Wang, H.~Li, T.-H. Chen, and A.~E. Hassan, ``A study of {C}/{C++} code weaknesses on {Stack} {Overflow},'' \emph{IEEE Transactions on Software Engineering}, vol.~48, no.~7, p. 2359–2375, 2022.

\bibitem{gong2024cosqa}
J.~Gong, Y.~Wu, L.~Liang, Z.~Zheng, and Y.~Wang, ``{CoSQA+}: Enhancing code search dataset with matching code,'' in \emph{arXiv:2406.11589}, 2024.

\bibitem{CHAN1996775sampling}
F.~Chan, T.~Chen, I.~Mak, and Y.~Yu, ``Proportional sampling strategy: guidelines for software testing practitioners,'' \emph{Information and Software Technology}, vol.~38, no.~12, p. 775–782, 1996.

\bibitem{tan2005datamining}
P.-N. Tan, M.~Steinbach, and V.~Kumar, \emph{Introduction to data mining}.\hskip 1em plus 0.5em minus 0.4em\relax Addison-Wesley Longman Publishing Co., Inc., 2005.

\bibitem{wang2020log}
B.~Wang, S.~Ying, and Z.~Yang, ``A log-based anomaly detection method with efficient neighbor searching and automatic {K} neighbor selection,'' \emph{Scientific Programming}, vol. 2020, p. 1–17, 2020.

\bibitem{allamaniscodedup2019}
M.~Allamanis, ``The adverse effects of code duplication in machine learning models of code,'' in \emph{Proceedings of the 2019 ACM SIGPLAN International Symposium on New Ideas, New Paradigms, and Reflections on Programming and Software}, 2019, p. 143–153.

\bibitem{lohr2021sampling}
S.~L. Lohr, \emph{Sampling: Design and Analysis}.\hskip 1em plus 0.5em minus 0.4em\relax Chapman and Hall/CRC, 2021.

\bibitem{tort1978sampling}
R.~D. Tortora, ``A note on sample size estimation for multinomial populations,'' \emph{The American Statistician}, vol.~32, no.~3, p. 100–102, 1978.

\bibitem{seaman1999quality}
C.~Seaman, ``Qualitative methods in empirical studies of software engineering,'' \emph{IEEE Transactions on Software Engineering}, vol.~25, no.~4, p. 557–572, 1999.

\bibitem{saldana2015coding}
J.~Saldana, \emph{The Coding Manual for Qualitative Researchers}.\hskip 1em plus 0.5em minus 0.4em\relax SAGE Publications, 2015.

\bibitem{barnad1947test}
G.~A. Barnard, ``Significance tests for 2×2 tables,'' \emph{Biometrika}, vol.~34, no. 1–2, p. 123–138, 1947.

\bibitem{bonferroni1936teoria}
C.~Bonferroni, \emph{Teoria statistica delle classi e calcolo delle probabilit{\`a}}.\hskip 1em plus 0.5em minus 0.4em\relax Seeber, 1936.

\bibitem{sharpe2015residual}
D.~Sharpe, ``Your chi-square test is statistically significant: now what?'' \emph{Practical assessment, research and evaluation}, vol.~20, no.~8, pp. 1--10, 2015.

\bibitem{Macke2021lineage}
S.~Macke, H.~Gong, D.~J.-L. Lee, A.~Head, D.~Xin, and A.~Parameswaran, ``Fine-grained lineage for safer notebook interactions,'' \emph{Proceedings of the VLDB Endowment}, vol.~14, no.~6, p. 1093–1101, 2021.

\bibitem{wei2024apimisuse}
M.~Wei, N.~S. Harzevili, Y.~Huang, J.~Yang, J.~Wang, and S.~Wang, ``Demystifying and detecting misuses of deep learning apis,'' in \emph{Proceedings of the IEEE/ACM 46th International Conference on Software Engineering}, 2024, p. 1–12.

\bibitem{Galappaththi2024apimisuse}
A.~Galappaththi, S.~Nadi, and C.~Treude, ``An empirical study of {API} misuses of data-centric libraries,'' in \emph{Proceedings of the 18th ACM/IEEE International Symposium on Empirical Software Engineering and Measurement}, 2024, p. 245–256.

\bibitem{Chattopadhyay20notebook}
S.~Chattopadhyay, I.~Prasad, A.~Z. Henley, A.~Sarma, and T.~Barik, ``What’s wrong with computational notebooks? pain points, needs, and design opportunities,'' in \emph{Proceedings of the 2020 CHI Conference on Human Factors in Computing Systems}, 2020, p. 1–12.

\end{thebibliography}
